\documentclass[a4paper,11pt]{article}
\pdfoutput=1 

\usepackage{jheppub} 

\usepackage[T1]{fontenc} 
\usepackage[]{amsmath}
\usepackage[]{graphicx}
\usepackage[mathscr]{eucal}
\usepackage{mathrsfs}
\usepackage[margin=10pt,font=small,labelfont=bf]{caption}
\usepackage{subcaption}
\usepackage{bm}
\usepackage{braket}

\newcommand{\rE}{\mathrm{E}}

\newcommand{\bu}{\mathbf{u}}
\newcommand{\bv}{\mathbf{v}}

\newcommand{\cA}{\mathcal{A}}
\newcommand{\cB}{\mathcal{B}}
\newcommand{\cC}{\mathcal{C}}
\newcommand{\cD}{\mathcal{D}}
\newcommand{\cS}{\mathcal{S}}

\newcommand{\tA}{\tilde{\cA}}
\newcommand{\tB}{\tilde{\cB}}
\newcommand{\tC}{\tilde{\cC}}
\newcommand{\tD}{\tilde{\cD}}

\newcommand{\pdif}[1]{\frac{\partial}{\partial #1}}
\newcommand{\qsz}{q^{S^z}}
\newcommand{\qmsz}{q^{-S^z}}
\DeclareMathOperator{\tr}{Tr}

\title{\boldmath On the Semi-Classical Limit of Scalar Products of the XXZ Spin Chain}

\author{Yunfeng Jiang,}
\author{Joren Brunekreef}

\affiliation[a]{Institut f{\"u}r Theoretische Physik,
ETH Z{\"u}rich,\\ Wolfgang Pauli Strasse 27,
CH-8093 Z{\"u}rich, Switzerland}

\emailAdd{jiangyf2008@gmail.com}
\emailAdd{jorenb@gmail.com}

\abstract{We study the scalar products between Bethe states in the XXZ spin chain with anisotropy $|\Delta|>1$ in the semi-classical limit where the length of the spin chain and the number of magnons tend to infinity with their ratio kept finite and fixed. Our method is a natural yet non-trivial generalization of similar methods developed for the XXX spin chain. The final result can be written in a compact form as a contour integral in terms of Faddeev's quantum dilogarithm function, which in the isotropic limit reduces to the classical dilogarithm function.}

\begin{document}
\maketitle
\flushbottom

\section{Introduction}
\label{sec:intro}
The quantum Heisenberg spin chain is one of the most well-known integrable models with a long history of research since the birth of the Bethe ansatz \cite{Bethe:Anstaz}. This seemingly simple model has surprisingly deep and rich physical and mathematical structures hidden under the surface and has triggered many important developments in the theory of quantum integrability for decades, see for example \cite{baxter2007exactly,Korepin:Book,jimbo1994algebraic,gaudin2014bethe} and references therein.\par

More recently, it has attracted considerable renewed interest from different branches of theoretical physics ranging from statistical mechanics out of equilibrium \cite{caux2013time,Ilievski:complete,eisert2015quantum} to AdS/CFT correspondence. In particular, it was found that the one-loop dilatation operator of the planar $\mathcal{N}=4$ Super-Yang-Mills ($\mathcal{N}=4$ SYM) theory coincides with the Hamiltonian of the Heisenberg XXX spin chain \cite{Minahan:2002ve,Beisert:2003yb}. This observation plays a crucial role in solving the theory exactly \cite{BigReview} since one can apply the powerful tools from integrability. On the other hand, in the quest of a better understanding of $\mathcal{N}=4$ SYM, one poses new questions on different aspects of the Heisenberg spin chain that were not emphasized in previous studies.\par

One such example is the so-called semi-classical limit of the Heisenberg XXX spin chain. This is the limit where the length of the spin chain $L$ and the number of magnons $N$ are large, but with their ratio $N/L=\alpha$ fixed and finite. In this limit, the solutions of the Bethe ansatz equations, or Bethe roots, condensate and form macroscopic cuts in the complex plane. This limit was first studied by Sutherland \cite{Sutherland} in the condensed matter physics literature. In the context of AdS/CFT, this limit is of great interest since the macroscopic cuts formed by Bethe roots are identified with the branch cuts of finite gap solutions of the classical string sigma model \cite{Beisert:2003xu,Kazakov:2004qf}, which relates integrable structures on both sides of the duality. The spectral problem in the semi-classical limit simplifies dramatically and can be formulated in an elegant way in terms of algebraic curves \cite{Kazakov:2004qf,Beisert:2005bm}.

It was first proposed in \cite{Okuyama:2004bd,Roiban:2004va} that the structure constants of $\mathcal{N}=4$ SYM theory can be computed in terms of scalar products of spin chains. This idea was further extended and elaborated systematically in \cite{Escobedo:2010xs}. The semiclassical limit of the structure constant was first studied in \cite{Gromov:2011jh} for a specific type of three-point functions (BSP-BPS-non BPS). Shortly after, a much more compact determinant formula for the structure constants was proposed in \cite{Foda:2011rr}. Based on the determinant formula, the semiclassical limit of more general three-point functions (three non-BPS operators) can be taken \cite{Kostov:3ptPRL}. Studies of three-point functions in the semiclassical limit show that structure constants in this limit also simplify significantly and can be expressed in a compact form in terms of contour integrals of dilogarithm functions. This same structure was obtained from weak \cite{Gromov:2011jh,Kostov:3ptPRL,Kostov:3ptLong,Kostov-Matsuo,Jiang:Fixing,Bettelheim:2014gma} and strong \cite{Kazama:2011cp,Kazama:2012is,Kazama:2013qsa,Kazama:2016cfl} coupling by rather different methods. Very recently, partial results of the same structure were derived at any coupling \cite{Basso:2015zoa,Jiang:2016ulr} by the clustering method. On the weak coupling side, the semi-classical limit of the structure constant can be obtained from taking the semi-classical limit of the scalar product of the type on-shell/off-shell, which allows a determinant representation \cite{Slavnov:determinant} called the Slavnov determinant.\par

At the same time, scalar products of Bethe states are of great importance in integrable spin chains of their own since they are fundamental building blocks of physical observables such as form factors and correlation functions of local spin operators \cite{KitanineFF,Kitanine:Review}. Therefore, the semi-classical limit of scalar products of the Heisenberg spin chain can also be regarded as a well-posed pure spin chain problem which is highly non-trivial and interesting in its own right. Various methods have been developed to take the semi-classical limit of scalar products of the Heisenberg XXX spin chain \cite{Gromov:2011jh,Kostov:3ptPRL,Kostov:3ptLong,Kostov-Matsuo,Jiang:Fixing,Bettelheim:2014gma} in the past few years. It is thus a natural question to ask whether some of these methods can be generalized to the XXZ spin chain, which is a $q$-deformed version of the XXX spin chain. And if so, how does the $q$-deformation affect the final result? The purpose of this paper is to investigate these interesting questions.\par

We find that indeed the methods \cite{Kostov:3ptLong,Kostov-Matsuo,Jiang:2016ulr} can be generalized to the XXZ case. In the XXZ spin chain, due to the presence of an anisotropy parameter $\Delta$, the structure of solutions of the Bethe ansatz equations is more complicated and depends on the range of $\Delta$. We find that the range $|\Delta|>1$ allows a most straightforward definition of the semi-classical limit that is similar to the XXX case. In this range, we take the semi-classical limit of the scalar product of the type on-shell/off-shell and obtain the following compact result
\begin{align}
\label{eq:semi}
&\text{XXX}:\qquad\log\langle\mathbf{v}|\mathbf{u}\rangle_{\text{XXX}}\sim \oint_{\mathcal{C}_{\mathbf{u}\cup\mathbf{v}}}\frac{du}{2\pi i}\int_0^{{g}_{\text{XXX}}(u)} \log(1-e^{i\mu})d\mu\\\nonumber
&\text{XXZ}:\qquad\log\langle\mathbf{v}|\mathbf{u}\rangle_{\text{XXZ}}\sim \oint_{\mathcal{C}_{\mathbf{u}\cup\mathbf{v}}}\frac{du}{2\pi i}\int_0^{{g}_{\text{XXZ}}(u)}  \log_q(1-e^{i\mu})d\mu
\end{align}
where the main difference is that the logarithm in the rational case is replaced by a $q$-analog of the logarithm in the trigonometric case. The $q$-analog is defined as
\begin{align}
\log_q(1-x)=-\sum_{n=1}^\infty \frac{x^n}{[n]_q},\qquad [n]_q=\frac{q^n-q^{-n}}{q-q^{-1}}.
\end{align}
The functions $g(u)$ of the two cases are given by
\begin{align}
g_{\text{XXX}}(u)=&\,-\frac{L}{u}+G^{\text{XXX}}_{\mathbf{u}}(u)+G^{\text{XXX}}_{\mathbf{v}}(u),\\\nonumber
g_{\text{XXZ}}(u)=&\,-\frac{L}{\tanh\gamma u}+G^{\text{XXZ}}_{\mathbf{u}}(u)+G^{\text{XXZ}}_{\mathbf{v}}(u).
\end{align}
where the resolvents $G_{\mathbf{u}}(u)$ for the two cases are
\begin{align}
G^{\text{XXX}}_{\mathbf{u}}(u)=\int_{A_{\mathbf{u}}}\frac{\rho_{\mathbf{v}}(v)}{u-v}dv,\qquad
G^{\text{XXZ}}_{\mathbf{u}}(u)=\int_{A_{\mathbf{u}}}\frac{\rho_{\mathbf{v}}(v)}{\tanh\gamma(u-v)}dv
\end{align}
Here $A_{\mathbf{v}}$ denotes the cut on which the Bethe roots are distributed and $\rho_{\mathbf{u}}(u)$ is the density of Bethe roots on the cut.
The full expression can be found in section\,\ref{sec:semi-classical-scalar-products}.\par

Interestingly, let us note that the semi-classical limit (\ref{eq:semi}) of the XXZ spin chain can actually be written in terms of Faddeev's quantum dilogarithm function $\Phi_b(z)$\footnote{We thank Ivan Kostov for pointing out this fact to us.} \cite{Faddeev:1993rs} as
\begin{align}
\log\langle\mathbf{v}|\mathbf{u}\rangle_{\text{XXZ}}\sim \oint_{\mathcal{C}_{\mathbf{u}\cup\mathbf{v}}}\frac{du}{2\pi i}\log\Phi_{\sqrt{\phi}}\left(  g_{\text{XXZ}}(u)+\pi\right)
\end{align}
where the anisotropy $\gamma=i\phi$, $\phi>0$. The definition of $\Phi_b(z)$ and its relation to the dilogarithm function are given in appendix\,\ref{sec:DL}.

The rest of this paper is structured as the follows. In section\,\ref{sec:A-functional} we briefly review the Algebraic Bethe Ansatz and introduce an important quantity called the $q$-deformed $\mathscr{A}$-functional by computing a special type of scalar product between an off-shell Bethe state and a vacuum descendant state. In section\,\ref{sec:Slavnov}, we rewrite the Slavnov determinant in terms of the $q$-deformed $\mathscr{A}$-functional following a similar method in \cite{Kostov-Matsuo}. In section\,\ref{sec:semi-classical-xxz} we define more carefully the semi-classical limit in the XXZ spin chain and in section\,\ref{sec:semi-classical-scalar-products} we take the semi-classical limit of the $q$-deformed $\mathscr{A}$-functional and the Slavnov determinant by generalizing the clustering method in \cite{Jiang:2016ulr} to the trigonometric case. We conclude in section\,\ref{sec:conclusion}. Appendix\,\ref{sec:app-comm-rels} and \ref{sec:app-large-rap} contain lists of commutation relations that are useful in the main text. In appendix\,\ref{app:numerics} we give more detail about the numerics on solving the Bethe ansatz equations of the XXZ spin chain. In appendix\,\ref{sec:DL} we give the definition of Faddeev's quantum dilogarithm function and its relation to the classical dilogarithm.\par

\textbf{Note added:} At the finishing stage of this paper, we became aware that the same problem was investigated by C. Babenko in \cite{Babenko} which has a significant overlap with the current paper.

\section{The $q$-deformed $\mathscr{A}$-functional}
\label{sec:A-functional}
In this section, we compute the scalar product of a generic off-shell Bethe state with a vacuum descendant state for the XXZ Heisenberg spin chain. The vacuum descendant state is defined by acting with generators of $U_q(\mathfrak{sl}(2))$ on the pseudovacuum state. In the XXX case, this scalar product gives rise to the so-called $\mathscr{A}$-functional, which plays an important role in computing the semi-classical limit of other scalar products \cite{Kostov:3ptLong}. Similarly, we define a $q$-deformed version of the $\mathscr{A}$-functional, which is subsequently used for obtaining scalar products of more general states.

Before we proceed to define the $\mathscr{A}$-functional, we briefly review the Algebraic Bethe Ansatz. This will also serve to set up our notations and conventions.

\subsection{Algebraic Bethe Ansatz of XXZ spin chain}
The Hamiltonian of the XXZ spin chain is given by
\begin{align}
    \label{eq:xxz-ham}
    H &= J \sum_{n=1}^L \left[ \sigma_n^x \sigma_{n+1}^x + \sigma_n^y \sigma_{n+1}^y + \Delta \left( \sigma_n^z \sigma_{n+1}^z -1\right) \right],
\end{align}
where $\Delta$ is the anisotropy. We impose a periodic boundary condition: $L+1\equiv 1$. For $\Delta = 1$, we recover the Hamiltonian of the XXX spin chain. The XXZ spin chain can be considered as a $q$-deformation of the XXX spin chain. To see this, we define the parameters $q$ and $\gamma$:
\begin{align}
    \frac{1}{2} \left(q+q^{-1}\right) \equiv \cos \gamma \equiv \Delta, \quad \quad q = e^{i \gamma}.
\end{align}
We then obtain the isotropic case $\Delta=1$ in the equivalent limits $q \to 1$ and $\gamma \to 0$. The XXZ spin chain is integrable and can be solved by the Algebraic Bethe Ansatz \cite{Faddeev:ABA}. The $q$-deformed Lax operator takes the form
\begin{equation}
    \hat{L}_{n,a}(u) = \left(
        \begin{array}{cc}
            \sinh \gamma\left(u + i S_n^z\right) & S_n^- \sin \gamma \\
            S_n^+ \sin \gamma & \sinh \gamma\left(u - i S_n^z\right)
        \end{array}
    \right).
\end{equation}
We can cast this into a more convenient form by defining the multiplicative spectral parameter $x = q^{-i u} = e^{\gamma u}$.\footnote{In what follows, we use a tilde to indicate that a function whose arguments are multiplicative parameters, and a hat for the additive parameters.} The Lax operator in terms of multiplicative parameters is then written as
\begin{align}
    \label{eq:aba-lax}
    \tilde{L}_{n,a}(x)&= \left(
    \begin{array}{cc}
      x q^{S_n^z}-x^{-1} q^{-S_n^z} & \left(q-q^{-1}\right) S_n^- \\
      \left(q-q^{-1}\right) S_n^+ & x q^{-S_n^z}-x^{-1} q^{S_n^z}
    \end{array} \right),
\end{align}
where we use the $q$-deformed operators
\begin{equation}
q^{\pm S_n^z} = \left( \begin{array} {cc} q^{\pm \frac{1}{2}} & 0 \\ 0 & q^{\mp \frac{1}{2}} \end{array} \right).
\end{equation}
The Lax operator satisfies the following $RLL$ relation:
\begin{equation}
    \label{eq:aba-fcr}
    R_{a,b}(x,y) L_{n,a}(x) L_{n,b}(y) = L_{n,b}(y) L_{n,a}(x) R_{a,b}(x,y)
\end{equation}
where the $R$-matrix takes the form
\begin{equation}
    \label{eq:aba-r-matrix}
    R_{a,b}(x,y) = \left(
        \begin{array}{cccc}
            a(x,y) & 0 & 0 & 0 \\
            0 & b(x,y) & c(x,y) & 0 \\
            0 & c(x,y) & b(x,y) & 0 \\
            0 & 0 & 0 & a(x,y)
        \end{array}
    \right),
\end{equation}
with the functions
\begin{equation}
    a(x,y) = q \frac{x}{y} - q^{-1} \frac{y}{x}, \quad \quad b(x,y) = \frac{x}{y}-\frac{y}{x}, \quad\quad c(x,y) = q-q^{-1}.
\end{equation}

The central quantity of the Algebraic Bethe Ansatz is the monodromy matrix $T_a(x)$ defined by:
\begin{equation}
    \label{eq:aba-monodromy}
    T_a(x) = L_{a,1}(x) L_{a,2}(x) \cdots L_{a,L}(x) = \left(
        \begin{array}{cc}
            \cA(x) & \cB(x) \\
            \cC(x) & \cD(x)
        \end{array}
    \right).
\end{equation}
The monodromy matrix satisfies the following $RTT$-relation:
\begin{equation}
    R_{a,b}(x,y) T_a(x) T_b(y) = T_b(y) T_a(x) R_{a,b}(x,y).
\end{equation}
This leads to the commutation relations between the elements of the monodromy matrix. We refer to appendix \ref{sec:app-comm-rels} for a list of these relations. From the monodromy matrix, we define the \emph{transfer matrix} $t(x) \equiv \tr_a T_a(x)$ which generates all the conserved charges of the system.

In order to construct the eigenstates of the transfer matrix, we start with the reference state $\ket{\Omega} \equiv \ket{\uparrow^L}$, with all spins pointing up. This is an eigenstate of the operators $\cA$ and $\cD$:
\begin{align}
    &\cA(x) \ket{\Omega} = a(x) \ket{\Omega}, \quad \quad \cD(x) \ket{\Omega} = d(x) \ket{\Omega}, \\
    &a(x) = \left(x q^{\frac{1}{2}}-x^{-1} q^{-\frac{1}{2}}\right)^L, \quad \quad d(x) = \left(x q^{-\frac{1}{2}}-x^{-1} q^{\frac{1}{2}} \right)^L.
\end{align}
We construct further eigenstates of the transfer matrix by acting the operator $\cB$ on the reference state:
\begin{equation}
    \ket{\mathbf{x}} = \prod_{i=1}^N \cB(x_i) \ket{\Omega}.
\end{equation}
Requiring that such a state is indeed an eigenstate leads to the Bethe equations for the rapidities $\mathbf{x}$. In terms of the additive spectral parameters $\mathbf{u}$, the Bethe equations read
\begin{equation}
    \label{eq:aba-bae}
    \left(\frac{\sinh{\gamma \left(u_j+\frac{i}{2}\right)}}{\sinh{\gamma \left(u_j-\frac{i}{2}\right)}}\right)^L = \prod_{k \neq j}^N \frac{\sinh{\gamma\left(u_j-u_k+i\right)}}{\sinh{\gamma \left(u_j-u_k-i\right)}}, \quad \quad j = 1, \cdots N.
\end{equation}

\subsection{Scalar products of XXZ spin chain}
\label{sec:afunc-scalar-products}
The scalar products we will compute are of the form
\begin{equation}
    \tilde{A}_N(\mathbf{x}) \equiv \bra{\Omega} \left(\mathcal{S}_q^+\right)^N \prod_{i=1}^N \mathcal{B}(x_i) \ket{\Omega}.
\end{equation}
where we use the multiplicative spectral parameters $x_i=e^{\gamma u_i}$. We call the state $\langle\Omega|(\mathcal{S}_q^+)^N$ and its dual $(\mathcal{S}_q^-)^N|\Omega\rangle$ the \emph{vacuum descendant states}. The operators $\mathcal{S}_q^\pm$ are defined by
\begin{equation}
    \mathcal{S}_q^\pm = \sum_{i=1}^L q^{S_1^z} \otimes \cdots \otimes q^{S_{i-1}^z} \otimes S_i^\pm \otimes q^{-S_{i+1}^z} \otimes \cdots \otimes q^{-S_L^z},
\end{equation}
where $L$ is the length of the spin chain. These operators together with $S_q^z=q^{\sum_{i=1}^L S_i^z}$ generate the quantum group $U_q(\mathfrak{sl}_2)$, reflecting the deformed symmetry of the model \footnote{In fact, this point is a bit more subtle due to boundary conditions. The Hamiltonian with periodic boundary condition does not commute with all the generators of the $U_q(\mathfrak{sl}_2)$ algebra. Only a specific choice of boundary condition gives Hamiltonian which is $U_q(\mathfrak{sl}_2)$ invariant. For a detailed discussion of this point, we refer to \cite{Pasquier:1989kd}.}. The commutation relations of the generators of $U_q(\mathfrak{sl}_2)$ with the elements of the monodromy matrix can be found by performing a large-rapidity expansion of the commutation relations obtained from the $RTT$-relation. In appendix \ref{sec:app-large-rap}, we show how to carry out this expansion and provide a list of the resulting relations.

By employing these commutation relations, it can be shown that for general $N$, the scalar product $\tilde{A}_N(\mathbf{x})$ takes the following form:
\begin{align}
    \label{eq:scalar-product-general}
    \tilde{A}_N(\mathbf{x}) &= (-1)^N q^{-(L+N-1)N/2} \left[N\right]_{q}! \prod_{i=1}^N \left(x^{-1}_i d(x_i)\right) \\
    & \quad \times \sum_{\alpha \cup \bar{\alpha}=\mathbf{x}} \left(-\frac{1}{q^{N-1}}\right)^{\left|\alpha\right|} \prod_{x_i \in \alpha, x_j \in \bar{\alpha}} \frac{q \frac{x_i}{x_j}-q^{-1}\frac{x_j}{x_i}}{\frac{x_i}{x_j}-\frac{x_j}{x_i}} \prod_{x_i \in \alpha} \left(q e^{i \tilde{p}(x_i)} \right)^L, \nonumber
\end{align}
where
\begin{gather*}
    e^{i L \tilde{p}(x)} = \frac{a(x)}{d(x)}, \\
    \left[N\right]_q! = \left[N\right]_q \times \left[N-1\right]_q \times \cdots \times \left[1\right]_q, \quad \quad \left[N\right]_q = \frac{q^{N}-q^{-N}}{q-q^{-1}}.
\end{gather*}
We now group parts of this expression together:
\begin{align}
    \tilde{K}_N(\mathbf{x}) &\equiv (-1)^N q^{-(L+N-1)N/2} \left[N\right]_{q}! \prod_{i=1}^N x^{-1}_i \\
    \tilde{\chi}(x) &\equiv  q^{L-N+1} e^{i L \tilde{p}(x)} \\
    \tilde{A}_N(\mathbf{x}) &= \tilde{K}_N(\mathbf{x}) \times \sum_{\alpha \cup \bar{\alpha}=\mathbf{x}} \left(-1\right)^{|\alpha|} \prod_{x_i \in \alpha} \tilde{\chi}(x_i) \prod_{\substack{x_i \in \alpha \\ x_j \in \bar{\alpha}}} \frac{q \frac{x_i}{x_j}-q^{-1}\frac{x_j}{x_i}}{\frac{x_i}{x_j}-\frac{x_j}{x_i}}.
\end{align}
In terms of additive spectral parameters $u_i$, we have
\begin{align}
    \label{eq:sp-additive-1}
    \hat{K}_N(\mathbf{u}) &\equiv (-1)^N q^{-(L+N-1)N/2} \left[N\right]_{q}! \prod_{i=1}^N e^{-\gamma u_i} \\
    \label{eq:sp-additive-2}
    \hat{\chi}(u) &\equiv  q^{L-N+1} e^{i L \hat{p}(u)} \\
    \label{eq:sp-additive-3}
    \hat{A}_N(\mathbf{u}) &= \hat{K}_N(\mathbf{u}) \times \sum_{\alpha \cup \bar{\alpha}=\mathbf{u}} \left(-1\right)^{|\alpha|} \prod_{u_i \in \alpha} \hat{\chi}(u_i) \prod_{\substack{u_i \in \alpha \\ u_j \in \bar{\alpha}}} \frac{\sinh \gamma(u_i-u_j+i)}{\sinh \gamma(u_i-u_j)}.
\end{align}

In the limit $q\to 1$, the expressions \eqref{eq:sp-additive-1} to \eqref{eq:sp-additive-3} reduce to the $\mathscr{A}$-functional for the XXX model (in the process, the trigonometric functions are replaced by their rational counterparts). Therefore, we take \eqref{eq:sp-additive-3} as the base for the $q$-deformed $\mathscr{A}$-functional, denoted by $\mathscr{A}^q_{\mathbf{u}}\left[\chi\right]$. For notational simplicity, we rescale the variables inside the function by $u_j\to u_j/\gamma$, and replace the $i$ by the parameter $\epsilon = i \gamma$. Our final definition of the $\mathscr{A}$-functional then reads
\begin{equation}
    \label{eq:afunc}
    \mathscr{A}^q_{\bu}\left[\chi\right] = \sum_{\alpha \cup \bar{\alpha}=\mathbf{u}} \left(-1\right)^{|\alpha|} \prod_{u_i \in \alpha} \chi(u_i) \prod_{\substack{u_i \in \alpha \\ u_j \in \bar{\alpha}}} \frac{\sinh(u_i-u_j+\epsilon)}{\sinh(u_i-u_j)}.
\end{equation}
The $q$-deformed $\mathscr{A}$-functional can be written equivalently as
\begin{equation}
    \label{eq:q-deformed-a-func}
    \mathscr{A}^q_{\mathbf{u}} \left[\chi \right] = \frac{1}{\Delta_{\mathbf{u}}} \prod_{j=1}^N \left(1-\chi(u_j) e^{\epsilon \partial/\partial u_j} \right) \Delta_{\mathbf{u}},
\end{equation}
where $e^{\epsilon \partial/\partial u}$ is the shift operator $e^{\epsilon \partial/\partial u} f(u) = f(u+\epsilon)$ and $\Delta_{\mathbf{u}}$ is the trigonometric Vandermonde determinant
\begin{equation}
    \label{eq:vandermonde}
    \Delta_{\mathbf{u}} = \prod_{j < k} \sinh(u_j-u_k).
\end{equation}
Finally, let us mention that the $\mathscr{A}$-functional is related to the so-called domain wall partition function of the 6-vertex model and various determinant formula are known. In the $q$-deformed case, a trigonometric version of the Kostov-Matsuo determinant formula \cite{Kostov-Matsuo} and variations have been derived in \cite{Foda:2012partial,Foda:2012variation}. However, in taking the semiclassical limit we are \emph{not} using any of the determinant formula. For our purpose, it is more useful to use the sum-over-partition formula (\ref{eq:q-deformed-a-func}) which can be converted to a multiple integral representation similar to the one in \cite{Jiang:2016ulr}.

\section{Slavnov determinant as an $\mathscr{A}$-functional}
\label{sec:Slavnov}
In this section, we will show that we can express the scalar product of an off-shell Bethe state with an on-shell Bethe state in terms of the $\mathscr{A}$-functional. We largely follow the derivations presented in \cite{Slavnov:determinant}, encountering some complications due to the rational functions being replaced by their trigonometric counterparts.

\subsection{Factorizing the Slavnov determinant}
It was shown by Slavnov \cite{Slavnov:determinant} that we can express the scalar product of a generic off-shell Bethe state with an on-shell state in terms of a determinant. Let the set of rapidities $\mathbf{u}$ be on-shell, i.e. satisfy the Bethe ansatz equations, and $\mathbf{v}$ be arbitrary of equal cardinality $N$. The scalar product is then given by
\begin{equation}
    \label{eq:off-on-shell-scalar-product}
    \braket{\bv | \bu} = \prod_{i=1}^N a(v_i) d(u_i) \mathscr{S}_{\bu,\bv},
\end{equation}
where $\mathscr{S}_{\bu,\bv}$ is the \emph{Slavnov determinant},
\begin{equation}
    \label{eq:slavnov-determinant}
    \mathscr{S}_{\bu,\bv} = \frac{1}{\prod_{j=1}^N a(v_j)} \frac{ \det_{j,k}  \pdif{u_j} T_{\bu}(v_k)}{\det_{j,k} \frac{1}{\sinh(u_j-v_k)}}.
\end{equation}
The function $T_{\bu}(v)$ is the eigenvalue of the transfer matrix:
\begin{equation}
    \label{eq:transfer-matrix}
    T_{\bu}(v)= a(v) \frac{Q_{\bu}(v-\epsilon)}{Q_{\bu}(v)} + d(v) \frac{ Q_{\bu}(v+\epsilon)}{Q_{\bu}(v)},
\end{equation}
where $Q_{\mathbf{u}}(v)$ is the Baxter $Q$-function defined by
\begin{align}
Q_{\mathbf{u}}(v)\equiv \prod_{k=1}^N\sinh(v-u_k).
\end{align}
Taking derivative with respect to one of the rapidities, we find
\begin{align}
-\frac{\partial}{\partial u_k}T_{\mathbf{u}}(v)=a(v)\frac{Q_{\mathbf{u}}(v-\epsilon)}{Q_{\mathbf{u}}(v)}\Omega(u_k,v),
\end{align}
where we defined the Slavnov kernel as
\begin{align}
    \Omega(u,v)&= t(u-v)-t(v-u)\,\frac{d(v)}{a(v)}\frac{Q_{\mathbf{u}}(v+\epsilon)}{Q_{\mathbf{u}}(v-\epsilon)}\\\nonumber
    &\equiv t(u-v)-t(v-u)\,e^{2ip_{\mathbf{u}}(v)}.
\end{align}
Here we use the same definition for the pseudo-momentum $e^{2ip_{\mathbf{u}}(v)}$ as presented in \cite{Kostov:3ptLong} and the function $t(u)$ is given by
\begin{align}
\label{eq:tu}
t(u)=\frac{\sinh\epsilon}{\sinh (u)\sinh(u+\epsilon)}=\frac{1}{\tanh(u)}-\frac{1}{\tanh(u+\epsilon)}.
\end{align}
Using the following identity
\begin{align}
\prod_{k=1}^N\frac{Q_{\mathbf{u}}(v_k-\epsilon)}{Q_{\mathbf{u}}(v_k)}\det_{j,k}\frac{1}{\sinh(u_j-v_k+\epsilon)}=\det_{j,k}\frac{1}{\sinh(u_j-v_k)},
\end{align}
we can write the Slavnov determinant simply as
\begin{align}
    \label{eq:slavnov-det-simple}
\mathscr{S}_{\mathbf{u},\mathbf{v}}=\frac{\det_{j,k}\Omega(u_j,v_k)}{\det_{j,k}\frac{1}{\sinh(u_j-v_k+\epsilon)}}.
\end{align}
One crucial step in computing the semi-classical limit of the scalar product is to write the Slavnov determinant in a factorized form of two $\mathscr{A}$-functionals. In the rational case, it is important that the function $t(u)$ can be written in the form $f(u) - f(u+\epsilon)$, for some function $f$. There it takes the following form:
\begin{align}
    t_{\text{XXX}}(u) &= \frac{\epsilon}{u(u+\epsilon)} = \frac{1}{u} - \frac{1}{u+\epsilon}.
\end{align}
For the XXZ spin chain, $t(u)$ is given by trigonometric functions. Nevertheless, from \eqref{eq:tu}, we see that it can still be rewritten in such a `shifted difference' form. We can thus write the $q$-deformed Slavnov kernel as
\begin{align}
\Omega(u,v)=\left(1-e^{2ip_{\mathbf{u}}(v)} e^{\epsilon\,\partial/\partial v}  \right)\left(e^{-\epsilon\,\partial/\partial u}-1\right)\coth(u-v+\epsilon).
\end{align}
Therefore, the numerator of \eqref{eq:slavnov-det-simple} can be written as
\begin{align}
\det_{j,k}\Omega(u_j,v_k)=&\,\prod_{i=1}^N\left(1-e^{2ip_{\mathbf{u}}(v_i)} e^{\epsilon\,\partial/\partial v_i}  \right)
\prod_{i=1}^N\left(e^{-\epsilon\,\partial/\partial u_i}-1\right)\det_{j,k}\frac{1}{\tanh(u_j-v_k+\epsilon)}\\\nonumber
=&\,\prod_{i=1}^N\left(1-e^{2ip_{\mathbf{u}}(v_i)} e^{\epsilon\,\partial/\partial v_i}  \right)
\prod_{i=1}^N\left(e^{-\epsilon\,\partial/\partial u_i}-1\right) \\
& \quad \quad \cdot \cosh\left(\sum_{j=1}^N \left(u_j-v_j\right) +N \epsilon\right) \det_{j,k}\frac{1}{\sinh(u_j-v_k+\epsilon)},
\end{align}
where we used the fact that
\begin{align}
    \det_{j,k} \frac{1}{\tanh\left(u_j-v_k+\epsilon\right)} = \cosh \left(\sum^N_{j=1}\left(u_j-v_j+\epsilon\right)\right) \det_{j,k}\frac{1}{\sinh\left(u_j-v_k+\epsilon\right) }.
\end{align}
In order to proceed, we need to move the hyperbolic cosine term to the left through all the shift operators. For convenience, we define
\begin{equation}
    \Xi_{\mathbf{u},\mathbf{v}}\equiv\sum_{j=1}^N \left(u_j-v_j + \epsilon\right), \quad \quad C_{\mathbf{u},\mathbf{v}} \equiv \cos\left(\Xi_{\mathbf{u},\mathbf{v}}\right).
\end{equation}
The action of the following two operators on $C_{\mathbf{u},\mathbf{v}}$ is fairly simple:
\begin{align}
    \left(e^{-\epsilon \partial/\partial u_i}-1\right) C_{\mathbf{u},\mathbf{v}} &= C_{\mathbf{u},\mathbf{v}} \left(\zeta_{\mathbf{u},\mathbf{v}} e^{-\epsilon \partial/\partial u_i}-1\right), \\
    \left(1-e^{2i p_{\mathbf{u}}(v_i)} e^{\epsilon \partial/\partial v_i}\right) C_{\mathbf{u},\mathbf{v}}  &= C_{\mathbf{u},\mathbf{v}} \left(1-e^{2i p_{\mathbf{u}}(v_i)} \zeta_{\mathbf{u},\mathbf{v}} e^{\epsilon \partial/\partial v_i}\right),
\end{align}
where
\begin{align}
    \zeta_{\mathbf{u},\mathbf{v}} = \frac{\cos\left(\Xi_{\mathbf{u},\mathbf{v}} - \epsilon\right)}{\cos\left(\Xi_{\mathbf{u},\mathbf{v}}\right)}.
\end{align}
The Slavnov determinant can then be written as
\begin{align}
    \label{eq:slavnov-det-cuv}
    \frac{\mathscr{S}_{\mathbf{u},\mathbf{v}}}{C_{\mathbf{u},\mathbf{v}}}
=\frac{\prod_{i=1}^N\left(1-e^{2ip_{\mathbf{u}}(v_i)} \zeta_{\mathbf{u},\mathbf{v}} e^{\epsilon\,\partial/\partial v_i}  \right)
\prod_{i=1}^N\left(\zeta_{\mathbf{u},\mathbf{v}} e^{-\epsilon\,\partial/\partial u_i}-1\right)\det_{j,k}\frac{1}{\sinh(u_j-v_k+\epsilon)}}{\det_{j,k}\frac{1}{\sinh(u_j-v_k+\epsilon)}}.
\end{align}
Defining
\begin{equation}
    \Pi_{\bu,\bv} = \prod_{j=1}^N\prod_{k=1}^N \sinh(u_j-v_k+\epsilon),
\end{equation}
the denominator of the Slavnov formula \eqref{eq:slavnov-det-simple} can then be written as
\begin{equation}
    \label{eq:slavnov-det-denom}
    \det_{j,k} \frac{1}{\sinh(u_j-v_k+\epsilon)} = \frac{\Delta_{\bu} \Delta_{-\bv}}{\Pi_{\bu,\bv}}.
\end{equation}
We thus have
\begin{align}
    \frac{\mathscr{S}_{\mathbf{u},\mathbf{v}}}{C_{\mathbf{u},\mathbf{v}}}
=\frac{\Pi_{\bu,\bv}}{\Delta_{\bu} \Delta_{\bv}} \prod_{i=1}^N\left(1-e^{2ip_{\mathbf{u}}(v_i)} \zeta_{\mathbf{u},\mathbf{v}} e^{\epsilon\,\partial/\partial v_i}  \right) \prod_{i=1}^N\left(\zeta_{\mathbf{u},\mathbf{v}} e^{-\epsilon\,\partial/\partial u_i}-1\right)
\frac{\Delta_{\bu} \Delta_{\bv}}{\Pi_{\bu,\bv}}.
\end{align}
The $\Delta_{\bv}$ term can be moved to the left through all terms containing only shift operators $e^{-\epsilon \pdif{u_i}}$ without being affected. Similarly, $\Delta_{\bu}$ can be moved to the right through all terms containing only shift operators of the form $e^{\epsilon \pdif{v_i}}$. It remains to show how the shift operators act on $\Pi^{-1}_{\bu,\bv}$:
\begin{align}
\label{eq:movePi}
e^{-\epsilon\,\partial/\partial u_j}\Pi^{-1}_{\mathbf{u},\mathbf{v}}=&\,\rE_{\mathbf{v}}^+(u_j)\,\Pi^{-1}_{\mathbf{u},\mathbf{v}}e^{-\epsilon\,\partial/\partial u_j}, \\\nonumber
e^{\epsilon\,\partial/\partial v_j }\Pi^{-1}_{\mathbf{u},\mathbf{v}}=&\,\rE_{\mathbf{u}}^-(v_j)\,\Pi^{-1}_{\mathbf{u},\mathbf{v}}e^{\epsilon\,\partial/\partial v_j },
\end{align}
where
\begin{align}
\rE_{\mathbf{u}}^\pm(v)=\frac{Q_{\mathbf{u}}(v\pm\epsilon)}{Q_\mathbf{u}(v)},\qquad
\rE_{\mathbf{v}}^\pm(u)=\frac{Q_{\mathbf{v}}(u\pm\epsilon)}{Q_\mathbf{v}(u)}.
\end{align}
Now we can write the Slavnov determinant formula in the following factorized form
\begin{align}
\frac{\mathscr{S}_{\mathbf{u},\mathbf{v}}}{C_{\mathbf{u},\mathbf{v}}}=&\,(-1)^N
\frac{1}{\Delta_{\mathbf{v}}}\prod_{j=1}^N\left(1-e^{2ip_{\mathbf{u}}(v_j)}\,\zeta_{\mathbf{u},\mathbf{v}} \rE_{\mathbf{u}}^-(v_j) e^{\epsilon\,\partial/\partial v_j}  \right)\Delta_{\mathbf{v}}\\\nonumber
&\,\times\frac{1}{\Delta_{\mathbf{u}}}\prod_{j=1}^N\left(1-\zeta_{\mathbf{u},\mathbf{v}} \rE_{\mathbf{v}}^+(u_j)\,e^{-\epsilon\,\partial/\partial u_j}\right)\Delta_{\mathbf{u}}\cdot1
\end{align}
Let us then define two operators related to the $\mathscr{A}$-functional
\begin{align}
\hat{\mathscr{A}}_{\mathbf{u}}^\pm\left[ \chi \right]=\frac{1}{\Delta_{\mathbf{u}}}\prod_{j=1}^N\left(1-\chi(u_j)e^{\pm\epsilon\partial/\partial u_j}\right)\Delta_{\mathbf{u}}.
\end{align}
We can now write
\begin{align}
\label{eq:factorA}
\frac{\mathscr{S}_{\mathbf{u},\mathbf{v}}}{C_{\mathbf{u},\mathbf{v}}}=(-1)^N\,
\hat{\mathscr{A}}^+_{\mathbf{v}}\left[e^{2ip_{\mathbf{u}}}\, \zeta_{\mathbf{u},\mathbf{v}}\,\rE_{\mathbf{u}}^-\right]\cdot
\hat{\mathscr{A}}^-_{\mathbf{u}}\left[\zeta_{\mathbf{u},\mathbf{v}}\, \rE_{\mathbf{v}}^+\right]\cdot 1.
\end{align}
Note that the factorization is not complete since the first operator can act non-trivially on the second.

\subsection{A symmetric representation }
In this section, we show how to rewrite \eqref{eq:factorA} as a single $\mathscr{A}$-functional. It is more convenient to consider the inhomogeneous XXZ spin chain, with inhomogeneities $\bm{\theta} = \left\{\theta_1,\cdots,\theta_L\right\}$. Then we can write
\begin{equation}
    a(u)=Q_{\bm{\theta}}\left(u+\frac{\epsilon}{2}\right), \quad \quad d(u) = Q_{\bm{\theta}}\left(u-\frac{\epsilon}{2}\right).
\end{equation}
For convenience, we then define $z_j = \theta_j+\frac{\epsilon}{2}$, so that we have
\begin{equation}
    \frac{d(u)}{a(u)}= \frac{1}{\rE_{\mathbf{z}}^+(u)}.
\end{equation}
Furthermore, we rewrite the pseudo-momentum as follows:
\begin{align}
e^{2ip_{\mathbf{u}}(v)}=\frac{1}{\rE^+_{\mathbf{z}}(v)}\frac{\rE_{\mathbf{u}}^+(v)}{\rE_{\mathbf{u}}^-(v)}.
\end{align}
These two expressions then allow us to write
\begin{align}
\frac{\mathscr{S}_{\mathbf{u},\mathbf{v}}}{C_{\mathbf{u},\mathbf{v}}}=(-1)^N\,
\hat{\mathscr{A}}^+_{\mathbf{v}}\left[\zeta_{\mathbf{u},\mathbf{v}} \frac{\rE_{\mathbf{u}}^+}{\rE_{\mathbf{z}}^+} \right]\cdot
\hat{\mathscr{A}}^-_{\mathbf{u}}\left[\zeta_{\mathbf{u},\mathbf{v}} \rE_{\mathbf{v}}^+\right]\cdot1.
\end{align}
To proceed, we apply the following two identities for the $\mathscr{A}$-functional \cite{Kostov-Matsuo}:
\begin{align}
\label{eq:Apm}
\hat{\mathscr{A}}_{\mathbf{u}}^-\left[f\right]=\hat{\mathscr{A}}_{\mathbf{u}}^+\left[-\frac{\rE_{\mathbf{u}}^-}{\rE_{\mathbf{u}}^+} f \right],\qquad
\hat{\mathscr{A}}_{\mathbf{u}}^+\left[f\right]=\hat{\mathscr{A}}_{\mathbf{u}}^-\left[-\frac{\rE_{\mathbf{u}}^+}{\rE_{\mathbf{u}}^-} f \right]
\end{align}
Furthermore, from the Bethe Ansatz equation we obtain
\begin{align}
\label{eq:BAE}
-\frac{\rE_{\mathbf{u}}^+(u_j)}{\rE_{\mathbf{u}}^-(u_j)}=\frac{1}{\rE^+_\mathbf{z}(u_j)}.
\end{align}
Combining \eqref{eq:Apm} and \eqref{eq:BAE}, we can rewrite the second $\mathscr{A}$-functional
\begin{align}
    \label{eq:slavnov-reduced-symmetric}
\frac{\mathscr{S}_{\mathbf{u},\mathbf{v}}}{C_{\mathbf{u},\mathbf{v}}}=(-1)^N\,
\hat{\mathscr{A}}^+_{\mathbf{v}}\left[\zeta_{\mathbf{u},\mathbf{v}} \frac{\rE_{\mathbf{u}}^+}{\rE_{\mathbf{z}}^+}\right]\cdot
\hat{\mathscr{A}}^+_{\mathbf{u}}\left[\zeta_{\mathbf{u},\mathbf{v}} \frac{\rE_{\mathbf{v}}^+}{\rE_{\mathbf{z}}^+}\right]\cdot1,
\end{align}
giving us an expression which is completely symmetric in $\bu$ and $\bv$.

We now proceed to show that we can obtain another expression for the Slavnov determinant formula, with only one $\mathscr{A}$-functional. First, we define
\begin{equation}
    \Pi'_{\mathbf{u},\mathbf{v}} = \prod_{i,j} \sinh(u_i-v_j)
\end{equation}
and note that
\begin{align}
    \Delta_{\mathbf{u}\cup\mathbf{v}}=\Delta_{\mathbf{u}}\,\Delta_{\mathbf{v}}\,\Pi'_{\mathbf{u},\mathbf{v}}
\end{align}
Now from the definition of $\mathscr{A}$-functional, we can write
\begin{align}
\mathscr{A}_{\mathbf{u}\cup\mathbf{v}}\left[ f \right]=\frac{1}{\Delta_{\mathbf{u}}\Delta_{\mathbf{v}}\Pi'_{\mathbf{u},\mathbf{v}}}
\prod_{j=1}^N\left(1-f(u_j)e^{\epsilon\,\partial/\partial u_j}\right)
\prod_{j=1}^N\left(1-f(v_j)e^{\epsilon\,\partial/\partial v_j}\right)
\Delta_{\mathbf{u}}\Delta_{\mathbf{v}}\Pi'_{\mathbf{u},\mathbf{v}}.
\end{align}
We move the factor $\Pi'_{\mathbf{u},\mathbf{v}}$ through the two products of shift operators using (\ref{eq:movePi}) and obtain
\begin{align}
\mathscr{A}_{\mathbf{u}\cup\mathbf{v}}\left[ f \right]=&\,
\frac{1}{\Delta_{\mathbf{u}}}\prod_{j=1}^N\left(1-\rE_{\mathbf{v}}^+(u_j)f(u_j)e^{\epsilon\,\partial/\partial u_j}\right)\Delta_{\mathbf{u}}\\\nonumber
&\times\frac{1}{\Delta_{\mathbf{v}}}\prod_{j=1}^N\left(1-\rE_{\mathbf{u}}^+(v_j)f(v_j)e^{\epsilon\,\partial/\partial v_j}\right)\Delta_{\mathbf{v}}\cdot1\\\nonumber
=&\,\hat{\mathscr{A}}^+_{\mathbf{u}}\left[\rE_{\mathbf{v}}^+\,f\right]\cdot
\hat{\mathscr{A}}^+_{\mathbf{v}}\left[\rE_{\mathbf{u}}^+\,f\right]\cdot1.
\end{align}
Comparing this to \eqref{eq:slavnov-reduced-symmetric}, we see that we can indeed write the Slavnov determinant formula in the following simple form:
\begin{align}
    \label{eq:slavnov-single-afunc}
    \mathscr{S}_{\mathbf{u},\mathbf{v}}=(-1)^N\,C_{\mathbf{u},\mathbf{v}}\,\mathscr{A}_{\mathbf{u}\cup\mathbf{v}}\left[\frac{\zeta_{\mathbf{u},\mathbf{v}}}{\rE^+_{\mathbf{z}}}\right].
\end{align}
In \cite{Kostov-Matsuo}, the analogous expression was computed for the XXX spin chain. We see that the main difference between our result and the XXX case is the appearance of the terms $C_{\bu,\bv}$ and $\zeta_{\bu,\bv}$. Putting back the anisotropy $\gamma$ and $\epsilon=i\gamma$ in those two terms, we find
\begin{align}
    \label{eq:cos-parameters}
    C_{\mathbf{u},\mathbf{v}} &= \cosh \gamma \left(\sum_{j=1}^N \left(u_j-v_j\right)+N i \right), \\
    \zeta_{\mathbf{u},\mathbf{v}} &= \frac{\cosh \gamma  \left(\sum_{j=1}^N \left(u_j-v_j\right)+(N-1) i  \right)}{\cosh \gamma \left(\sum_{j=1}^N \left(u_j-v_j\right)+N i\right)} .
\end{align}
It is not hard to see that these factors tend to 1 in the limit $\gamma \to 0$, so the formula correctly reduces to its isotropic form.

\section{Semi-classical limit of XXZ spin chain}
\label{sec:semi-classical-xxz}
\begin{figure}[t]
    \centering
    \includegraphics[width=0.3\linewidth]{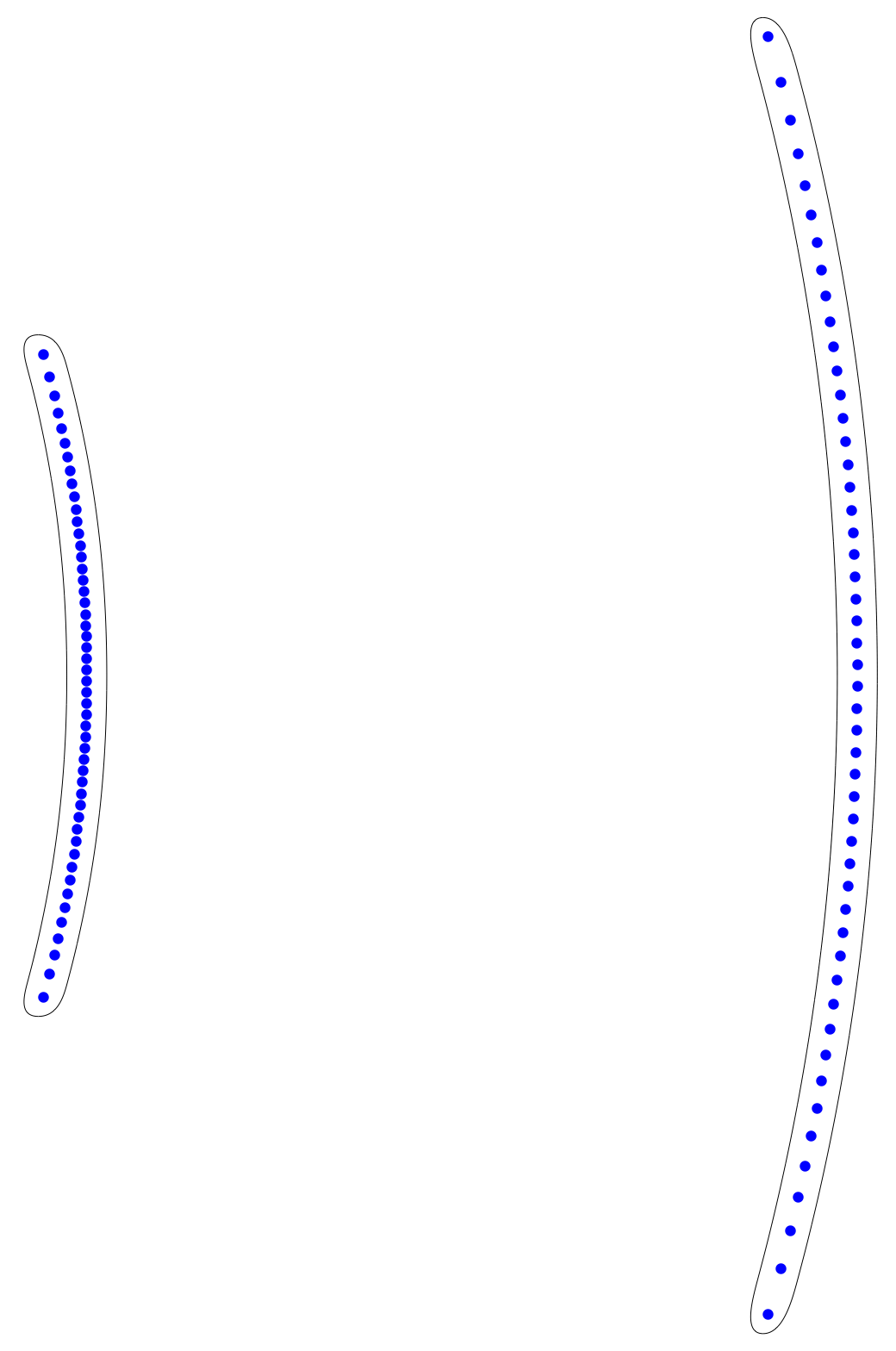}
    \caption{An example of the root distributions and corresponding contours for an \mbox{$L=1000, N=50$} spin chain. The two sets correspond to the solutions with mode number $n=1,2$.}
    \label{fig:contours-xxx}
\end{figure}

In this section we discuss the semi-classical limit of the XXZ spin chain. The semi-classical limit is defined as the limit where $N, L \to \infty$, while keeping their ratio $\alpha = N/L$ fixed. For the XXX model, the distribution of Bethe roots in this limit condenses into macroscopic cuts on the complex plane. An example of two root distributions with $L=1000$ and $N=50$ is given in figure \ref{fig:contours-xxx}. The fact that the distribution of Bethe roots forms macroscopic cuts enables us to describe physical quantities including the scalar products in terms of the density of Bethe roots $\rho(u)$ instead of individual Bethe roots, which in many cases leads to drastic simplifications.\par
In order to make a sensible generalization of the semi-classical limit techniques to the XXZ model, we should first make sure that we can still find solutions of the Bethe Ansatz equations that have similar distributions. To this end, we discuss some of the general aspects of the solutions of the XXZ Bethe Ansatz equations.

\subsection{XXZ root distributions}
\begin{figure}[t]
    \centering
    \includegraphics[width=0.7\linewidth]{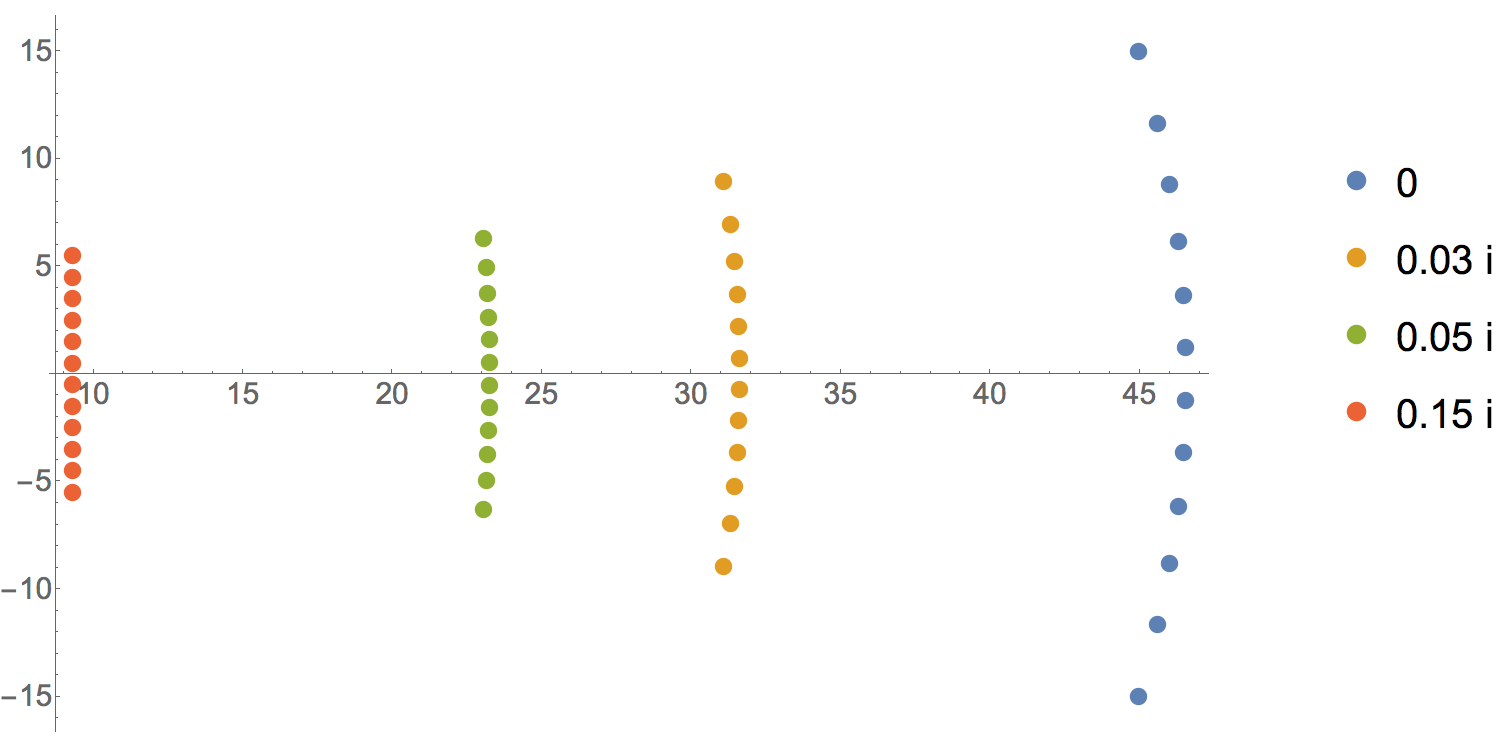}
    \caption{Root distributions for an $L=300, N=12$ spin chain. The rightmost set represents the isotropic case ($\gamma = 0$). The others represent the roots for chains with small anisotropies $\gamma=0.03i, \gamma=0.05i, \gamma=0.15i$, ordered from right to left. For $\gamma=0.15i$, the roots are already very close to the string distribution as defined in \eqref{eq:xxz-string-sol}, with $\lambda \approx 9.31$. }
    \label{fig:contours-xxz}
\end{figure}

We first recall the XXZ Bethe Ansatz equations:
\begin{equation}
    \label{eq:xxz-bethe-eq}
    \left(\frac{\sinh{\gamma \left(u_j+\frac{i}{2}\right)}}{\sinh{\gamma \left(u_j-\frac{i}{2}\right)}}\right)^L = \prod_{j \neq i}^N \frac{\sinh{\gamma\left(u_j-u_k+i\right)}}{\sinh{\gamma \left(u_j-u_k-i\right)}}, \quad \quad i = 1, \cdots N.
\end{equation}
The case of an anisotropy parameter $|\Delta|<1$ is known to give rise to complicated Bethe root distributions \cite{Takahashi:stacks}. However, for $|\Delta|> 1$, we can still find string solutions, similar to those found in the isotropic case. This region of $\Delta$ corresponds to a purely imaginary parameter $\gamma \equiv i \phi \equiv \cosh^{-1} \Delta, \, \phi > 0$. The string solutions can then be written \cite{Takahashi:stacks}:
\begin{equation}
    \label{eq:xxz-string-sol}
    u_j =  \lambda + \frac{1}{2} \left(M+1-2j\right)i,
\end{equation}
where $\lambda$ is a real number, $-\frac{\pi}{\phi} < \lambda < \frac{ \pi}{\phi}$. It is clear that all the trigonometric functions in the Bethe equations are then periodic along the direction of the real axis, so that adding or subtracting multiples of $\frac{2\pi}{\phi}$ generates equivalent sets of solutions. We present a sample of several Bethe root distributions with small anisotropy parameters in figure \ref{fig:contours-xxz}. We refer to appendix \ref{app:numerics} for more details on how these roots were obtained.

From the form of the string solutions given in \eqref{eq:xxz-string-sol}, we see that the majority of the roots will scale like $\left|u_j\right| \sim N$. In the semi-classical limit, we also have $N \sim L$, so the $u_j$ is of order $L$. We therefore can write
\begin{align}
\label{eq:largesin}
\sin\gamma(u_j-u_k+i)=\sin\gamma L(u'_j-u'_k+\epsilon')
\end{align}
where $\epsilon'=i/L$ is a small parameter. In fact, the parameter $\epsilon$ plays the role of Planck's constant in this context, since we obtain the semi-classical limit by letting it approach zero.\par

As a final remark of this section, we notice that expressions like (\ref{eq:largesin}) occur frequently in our derivations as well as in the Bethe equations and they typically take the form of a ratio
\begin{align}
\frac{\sinh\gamma(u-v+ia)}{\sinh\gamma(u-v+ib)}=\frac{\sinh\gamma'(u'-v'+a\epsilon)}{\sinh\gamma'(u'-v'+b\epsilon)},\qquad \gamma'=\gamma L.
\end{align}
For the rational case, the common factors $\gamma'$ in the numerator and denominator cancel each other. However, in the trigonometric case we do not have such cancellations. For $\gamma=i\phi$, $\phi\in\mathbb{R}$, if $|\gamma'|\to \infty$ the function $\sinh(\gamma'x)$ with $x\sim\mathcal{O}(1)$ is quickly oscillating. This makes the computations like solving Bethe equations tricky. Therefore, as a working assumption, we need to keep $\gamma'$ at some reasonably finite value. This means in practice if we take large $L$, we will take small $\gamma$.

\section{Semi-classical limit of scalar products}
\label{sec:semi-classical-scalar-products}
We have shown in section \ref{sec:Slavnov} that an arbitrary on-shell/off-shell scalar product can be written in terms of the $\mathscr{A}$-functional. Therefore, the problem of computing semi-classical limits of on-shell/off-shell scalar products in the XXZ chain reduces to obtaining the semi-classical limit of the $\mathscr{A}$-functional. In this section, we show in detail how to take the semi-classical limit of the $\mathscr{A}$-functional.

\subsection{The $\mathscr{A}$-functional as a grand partition function}
Recall the expression for the $q$-deformed $\mathscr{A}$-functional:
\begin{equation}
    \mathscr{A}^q_{\bu}\left[\chi\right] = \sum_{\alpha \cup \bar{\alpha} = \bu} \prod_{u_i \in \alpha}\left(- \chi(u_i) \right)\prod_{\substack{u_i \in \alpha \\ u_j \in \bar{\alpha}}} \frac{\sinh\left(u_i-u_j+\epsilon\right)}{\sinh \left(u_i-u_j\right)}.
\end{equation}
Using a similar method to \cite{Jiang:2016ulr}, we can rewrite the $\mathscr{A}$-functional in terms of a multiple contour integral
\begin{equation}
    \label{eq:afunc-contour}
    \mathscr{A}^q_{\bu}\left[\chi\right] = \sum_{n=0}^{\infty} \frac{1}{n!} F_n\left[\chi\right].
\end{equation}
where
\begin{equation}
    F_n\left[\chi\right] = \oint_{\mathcal{C}_{\bu}} \prod_{j=1}^n \left[\frac{d z_j}{2 \pi i} \right]\det_{j,k} \frac{1}{\sinh(z_j-z_k+\epsilon)} \prod_{j=1}^n \frac{Q_{\bu}(z_j+\epsilon)}{Q_{\bu}(z_j)} (- \chi(z_j)).
\end{equation}
The contour $\mathcal{C}_{\bu}$ is chosen so that it tightly encircles the Bethe roots $\bu$. To be more precise, all the integration variable $z_j$ share the same contour $\mathcal{C}_{\mathbf{u}}$ which encircles only the poles $z_j=u_k$ from the Baxter $Q$-functions and leaves the poles from the Cauchy determinant outside the contour. In this paper, we take $L$ large but finite, which means $\epsilon$ is small but not exactly zero so the aforementioned choice of contour is well-defined. The $n$-th term in the series then corresponds to the sum of all partitions with $n$ roots in the set $\alpha$, and the remaining $N-n$ roots in the set $\bar{\alpha}$. For $n > N$, the integrand is holomorphic within the region enclosed by the contour, so these terms give a zero contribution. We can define a slightly more general version of the $\mathscr{A}$-functional
\begin{equation}
    \mathscr{A}^{q,\kappa}_{\bu} = \sum_{n=0}^{\infty} \frac{\kappa^n}{n!} F_n.
\end{equation}
We rewrite the determinant in each $F_n$ as a sum over permutations:
\begin{equation}
    F_n = \sum_{\sigma \in S_n} (-1)^{\epsilon(\sigma)} \oint_{\cC_{\mathbf{u}}} \prod_{j=1}^n \left[\frac{d z_j}{2 \pi i} \frac{-\chi(z_j)}{\sinh\left(z_j-z_{\sigma(j)}+\epsilon\right)}\frac{ Q_{\bu}(z_j+\epsilon)}{Q_{\bu}(z_j)} \right].
\end{equation}
Then we define the function
\begin{equation}
    \rho(x,y) = \frac{1}{\sinh(x-y+\epsilon)}\left(\frac{Q_{\bu}(x+\epsilon)}{Q_{\bu}(x)}(- \chi(x)) \right),
\end{equation}
which allows us to write $F_n$ in the simpler form
\begin{equation}
    F_n=\sum_{\sigma \in S_n} (-1)^{\epsilon(\sigma)} \oint_{\cC_{\mathbf{u}}} \prod_{j=1}^n \frac{d z_j}{2 \pi i} \rho\left(z_j,z_{\sigma(j)}\right).
\end{equation}
The sum over permutations can then be transformed into a sum over conjugacy classes (see for example section 2.8 of \cite{feynman1998statistical} for a more detailed treatment). Every conjugacy class of the permutation group of order $n$ is characterized by a set of integers $\left\{C_{\ell}\right\}$ where $C_\ell$ denotes the number of conjugacy classes of length $\ell$. The integers $C_\ell$ satisfies the following constraint
\begin{equation}
    \sum_{\ell} \ell C_{\ell} = n.
\end{equation}
Therefore the sum over permutations in the expression for $F_n$ can be taken as a sum over integers satisfying this constraint. We denote such a sum with a prime, i.e. $\sum_{\left\{C_{\ell}\right\}}^{\prime}$. Defining
\begin{equation}
    \label{eq:fredholm-z}
    Z_{\ell} = \oint_{\cC_{\mathbf{u}}} \prod_{j=1}^{\ell} \left[ \frac{d z_j}{2 \pi i} \right] \rho(z_1, z_2) \rho(z_2, z_3) \cdots \rho(z_{l-1},z_{\ell}) \rho(z_{\ell},z_1).
\end{equation}
The term $F_n$ then takes the form
\begin{equation}
    \frac{F_n}{n!} = \sum_{\left\{C_{\ell}\right\}}' \prod_{\ell} \frac{(-1)^{(l-1)C_{\ell}} Z_{\ell}^{C_{\ell}}}{C_{\ell}! \ell^{C_{\ell}}}.
\end{equation}
This expression becomes highly non-trivial to write down for large $n$ due to the constraint. However, for the $\mathscr{A}$-functional we take the sum over all $n$. So the constraints disappear and the numbers $C_{\ell}$ can then take any non-negative integer values. We can now interchange the sum over $C_{\ell}$ with the product over $\ell$ and compute
\begin{align}
\label{eq:afunc-exponential}
\mathscr{A}_{\bu}^{q,\kappa} &= \sum_{n=0}^{\infty} \frac{\kappa^n F_n}{n!} = \sum_{\left\{C_{\ell}\right\}} \prod_{\ell} \frac{(-1)^{(\ell-1)C_{\ell}} \kappa^{\ell C_{\ell}} Z_{\ell}^{C_{\ell}}}{C_{\ell}! \ell^{C_{\ell}}} \nonumber \\
&= \sum_{\left\{C_{\ell} \right\}} \prod_{\ell} \frac{1}{C_{\ell}!} \left(- \frac{(-\kappa)^{\ell} Z_{\ell}}{\ell}\right)^{C_{\ell}} = \prod_{\ell} \sum_{\left\{C_{\ell}\right\}} \frac{1}{C_{\ell}!} \left(- \frac{(-\kappa)^{\ell} Z_{\ell}}{\ell}\right)^{C_{\ell}} \nonumber \\
&= \prod_{\ell} \exp \left(-\frac{(-\kappa)^{\ell} Z_{\ell}}{\ell}\right)
= \exp\left(-\sum_{\ell} Z_{\ell} \frac{(-\kappa)^{\ell}}{\ell} \right).
\end{align}
This expression is our starting point for obtaining the semi-classical limit.

\subsection{Semi-classical limit of the $\mathscr{A}$-functional}
\begin{figure}[h!]
    \centering
    \includegraphics[scale=0.6]{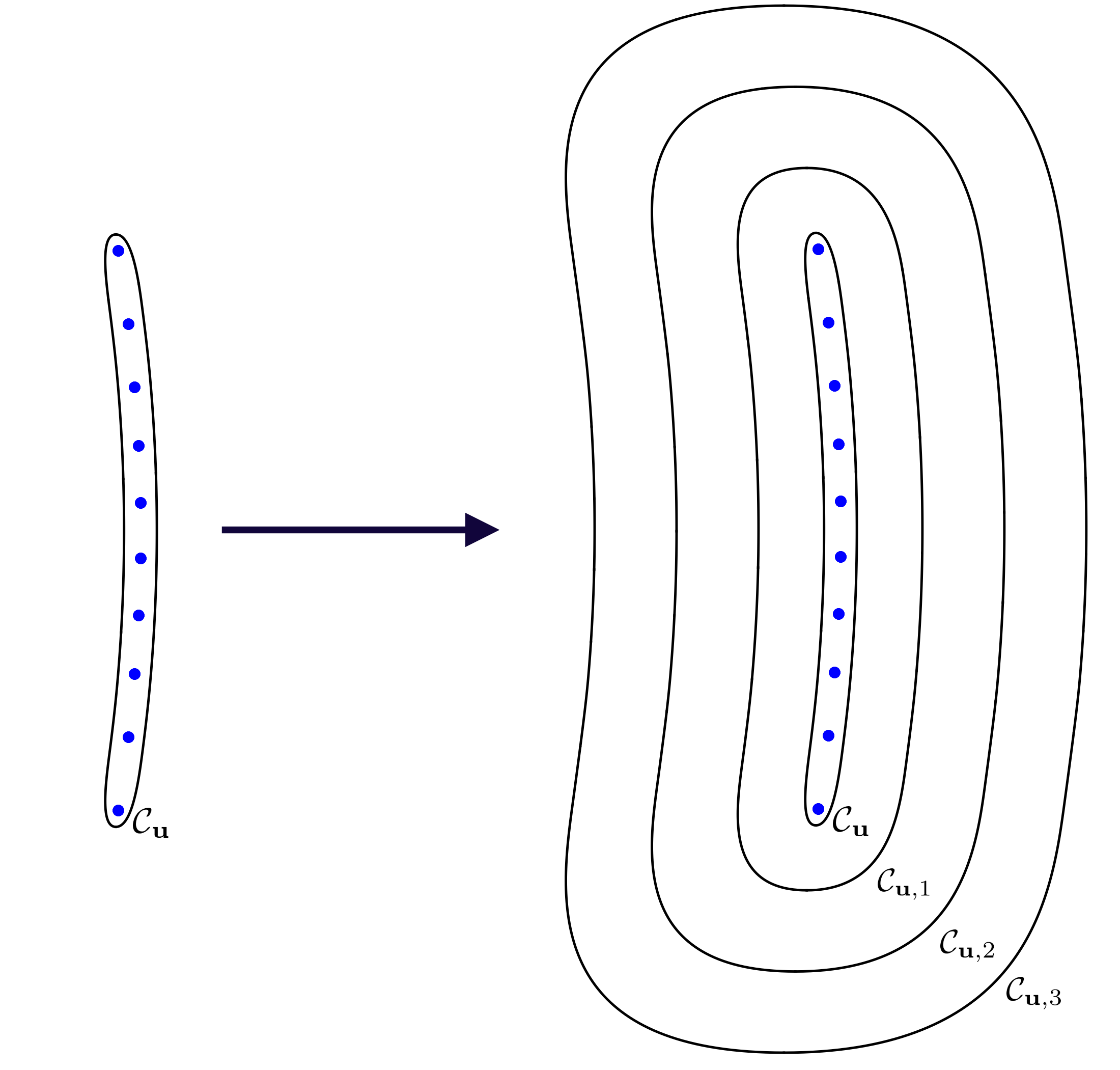}
    \caption{Deformation of the integration contours. The contour $\mathcal{C}_{\bu,k}$ corresponds to the contour for the integration variable $z_k$.}
    \label{fig:cont-def}
\end{figure}

We consider the multiple contour integrals $Z_{\ell}$ in \eqref{eq:fredholm-z} following the method given in \cite{Jiang:2016ulr}. The idea is to deform the multiple contours for every $z_j$ sequentially so that they are separated by at a distance larger than $\epsilon$, as is depicted in figure\,\ref{fig:cont-def}.
During the contour deformation, one picks up poles from the Cauchy determinant $z_j = z_k + \epsilon$, after which we find the simpler form
\begin{align}
\label{eq:semiZl}
    Z_{\ell} &= \oint_{\cC_{\mathbf{u}}} \prod_{j=1}^{\ell}\left[ \frac{d z_j}{2 \pi i} \frac{-G(z_j)}{\sinh\left(z_j-z_{j+1}+\epsilon\right)} \right] \\
&= (-1)^l \oint_{\cC_{\mathbf{u}}} \frac{d z}{2 \pi i} \frac{G(z) G(z+\epsilon) \cdots G(z+(\ell-1) \epsilon)}{\sinh \ell \epsilon},
\end{align}
where
\begin{equation}
    G(z) = \chi(z) \frac{Q_{\bu}(z+\epsilon)}{Q_{\bu}(z)}.
\end{equation}
We have seen in section \ref{sec:semi-classical-xxz} that $\epsilon$ is small compared to the variables $u_j$ (and therefore $z$) in the semi-classical limit. This allows us to make the approximation at leading order
\begin{equation}
    G(z+n \epsilon) \approx G(z),
\end{equation}
which we can plug back into the expression for $Z_{\ell}$. This can then be substituted into \eqref{eq:afunc-exponential}, which gives us (after setting $\kappa = 1$) the final expression for the $\mathscr{A}$-functional in the semi-classical limit:
\begin{equation}
    \label{eq:afunc-sc}
    \log \mathscr{A}_{\bu}^q[\chi] =- \oint_{\mathcal{C}_{\bu}} \frac{dz}{2 \pi i} \sum_{n=1}^{\infty} \frac{(G(z))^n}{n \sinh n \epsilon}.
\end{equation}
After the contour deformation, we now have only one contour integral in the final result (\ref{eq:afunc-sc}). This contour of this integral is chosen to be tightly encircling the support of the Bethe roots or the cut in semi-classical limit and it can be taken as the original contour $\mathcal{C}_{\mathbf{u}}$. Putting back the anisotropy explicitly and $\epsilon=i \gamma$, we can write the semi-classical limit of $\mathscr{A}_{\mathbf{u}}^q[\chi]$ as
\begin{align}
\label{eq:semiAXXZ}
\log\mathscr{A}_{\mathbf{u}}^q[\chi]\approx \oint_{\mathcal{C}_\mathbf{u}}\frac{dz}{2\pi}\sum_{n=1}^\infty\frac{(G(z))^n}{n\sin\gamma n}
=\frac{1}{\sin\gamma}\oint_{\mathcal{C}_{\mathbf{u}}}\frac{dz}{2\pi}\sum_{n=1}^\infty\frac{(G(z))^n}{n\,[n]_q}
\end{align}
where we have used the simple fact that
\begin{align}
[n]_q=\frac{q^n-q^{-n}}{q-q^{-1}}=\frac{\sin\gamma n}{\sin\gamma}
\end{align}
Comparing to the semi-classical limit of the $\mathscr{A}$-functional of the XXX spin chain \cite{Kostov:3ptLong}
\begin{align}
\label{semiAXXX}
\log\mathscr{A}_{\mathbf{u}}^{\text{XXX}}[\chi]\sim \oint_{\mathcal{C}_\mathbf{u}}\frac{dz}{2\pi}\sum_{n=1}^\infty\frac{(G(z))^n}{n^2}
\end{align}
we find that one of the factors $n$ in the infinite series of (\ref{semiAXXX}) is $q$-deformed while the other remains untouched. The interpretation is quite simple. In fact the two factors $n$ have different origins, one comes purely from combinatorics (\ref{eq:afunc-exponential}) which is the same for both XXX and XXZ spin chains while the other comes from (\ref{eq:semiZl}) which is model dependent.\par

In order to simplify our expression, we define formally a $q$-analog of the logarithm function
\begin{align}
\log_q(1-x)=-\sum_{n=1}^{\infty}\frac{x^n}{[n]_q}
\end{align}
which is a $q$-deformation of the usual expansion
\begin{align}
\log(1-x)=-\sum_{n=1}^{\infty}\frac{x^n}{n}.
\end{align}
Then we can write the semi-classical limit of the $\mathscr{A}$-functional as
\begin{align}
    \log\mathscr{A}_{\mathbf{u}}^q[\chi]\approx \frac{1}{\sin \gamma} \oint_{\mathcal{C}_{\bu}} \frac{du}{2 \pi i }\int_0^{g(u)}\log_q(1-e^{i\mu})d\mu,\qquad e^{ig(z)}=G(z).
\end{align}
Finally, it is interesting to note that our result (\ref{eq:semiAXXZ}) can also be written in terms of a function known as Faddeev's quantum dilogarithm $\Phi_b(z)$, which we define in (\ref{eq:QD}). In terms of this function, we can write
\begin{align}
\log\mathscr{A}_{\mathbf{u}}^q[\chi]=\oint_{\mathcal{C}_{\mathbf{u}}}\frac{dz}{2\pi i}\log\Phi_{\sqrt{\phi}}\left(g(z)+\pi\right)
\end{align}
Recalling that the semi-classical limit of the $\mathscr{A}$-functional for the XXX spin chain is given in terms of the dilogarithm function, our result for the XXZ spin chain can be seen as a sort of `quantization' of the XXX case.

\subsection{Semi-classical limit of Slavnov determinant}
Now it is straightforward to take the semi-classical limit of the Slavnov determinant $\mathscr{S}_{\mathbf{u},\mathbf{v}}$. From (\ref{eq:slavnov-single-afunc})
\begin{align}
\log\mathscr{S}_{\mathbf{u},\mathbf{v}}=i\pi N+\log C_{\mathbf{u},\mathbf{v}}+\log\mathscr{A}_{\mathbf{u}\cup\mathbf{v}}\left[{\zeta_{\mathbf{u},\mathbf{v}}}\,\frac{d(u)}{a(u)}\right]
\end{align}
In order to write down the semi-classical limit of the Slavnov determinant, it now remains to take the semi-classical limit of $C_{\bu,\bv}$ and
\begin{align}
G(z)=\zeta_{\mathbf{u},\mathbf{v}}\frac{d(z)}{a(z)}\frac{Q_{\mathbf{u}}(z+\epsilon)}{Q_{\mathbf{u}}(z)}\frac{Q_{\mathbf{v}}(z+\epsilon)}{Q_{\mathbf{v}}(z)}
\end{align}
For a proper string solution $\bu$, the sum $\sum_{\bu} u_j$ is real since the roots are distributed symmetrically about the real axis. For the off-shell rapidities $\bv$, we assume `reasonable' behavior, meaning that the rapidities condense on some cuts and the imaginary part of $\sum_{\bv} v_j$ is not too far from zero. First we consider the term $C_{\bu,\bv}$. Comparing to \eqref{eq:cos-parameters}, we find in the semi-classical limit
\begin{align}
    C_{\bu,\bv} =& \cosh \gamma \left( \sum_{j=1}^N \left(u_j-v_j\right)+N i \right) \\\nonumber
    \approx & \cosh \left(\gamma\int_{A_{\mathbf{u}}} du\,u\,\rho(u)-\gamma\int_{A_{\mathbf{v}}} dv\,v\,\rho(v) + i\gamma N \right).
\end{align}
The semi-classical limit of $G(z)$ can be taken easily as
\begin{align}
\log G(z)\approx -i \gamma\frac{L}{\tanh\gamma z}+i \gamma\int_{A_{\mathbf{u}}}\frac{\rho_{\mathbf{u}}(v)}{\tanh\gamma(z-v)}dv+i \gamma \int_{A_{\mathbf{v}}}\frac{\rho_{\mathbf{v}}(v)}{\tanh\gamma(z-v)}dv
\end{align}
where we have used the fact that $\log\zeta_{\mathbf{u},\mathbf{v}}$ is a small number of order 1. Here $\rho_{\mathbf{u}}(u)$ denotes the density of Bethe roots on the cut $\mathbf{u}$ and $A_{\mathbf{u}}$ denotes the cut on which the Bethe roots condensate. Then the semi-classical limit of the Slavnov determinant is given by
\begin{align}
    \log\mathscr{S}_{\mathbf{u},\mathbf{v}}=i\pi N + \log C_{\bu,\bv} +\frac{1}{\sin\gamma}\oint_{\mathcal{C}_{\bu\cup\mathbf{v}}}\frac{du}{2\pi i}\int_0^{g(u)}\log_q(1-e^{i\mu})d\mu.
\end{align}
Here the function $g(u)$ is given by
\begin{align}
g(u)=-\frac{\gamma L }{\tanh\gamma u}+G_{\mathbf{u}}(u)+G_{\mathbf{v}}(u)
\end{align}
where $G_{\mathbf{u}}(u)$ is the resolvent defined by
\begin{align}
G_{\mathbf{u}}(u)= \gamma \int_{A_{\mathbf{u}}}\frac{\rho_{\mathbf{u}}(v)}{\tanh\gamma(u-v)}dv.
\end{align}

\section{Conclusions and outlook}
\label{sec:conclusion}
In this paper, we investigated scalar products of Bethe states of the type on-shell/off-shell in the semi-classical limit for the XXZ spin chain with anisotropy $|\Delta|>1$. We define a quantity called the $q$-deformed $\mathscr{A}$-functional from the scalar product between a generic off-shell Bethe state and a vacuum descendant state. We then show the scalar product of the type on-shell/off-shell can be written in terms of the $\mathscr{A}$-functional. By generalizing the techniques of the XXX spin chain to the trigonometric case, we are able to take the semi-classical limit of the $q$-deformed $\mathscr{A}$-functional and the on-shell/off-shell scalar product. The final result can be written in terms of Faddeev's quantum dilogarithm function, which is a natural $q$-deformation of the classical dilogarithm function.\par

The original motivation of studying the semi-classical limit stems from AdS/CFT correspondence. It is known that the all loop $S$-matrix underlying $\mathcal{N}=4$ SYM allows a quantum deformation \cite{Beisert:2008tw} and there are also proposals for $q$-deformation of superstring theory on AdS$_5\times S^5$ \cite{Delduc:2013qra,Delduc:2014kha,Hollowood:2014qma}. It would be interesting to see whether our results can be applied to the context of $q$-deformations of AdS/CFT correspondence.\par

It would also be interesting to obtain systematically the $1/L$ corrections for the scalar products. This might be done by generalizing the methods in \cite{Bettelheim:Semi} or applying a Wigner-Kirkwood method in the Fermi gas approach \cite{Marino:2011eh}. Finally, we only analyzed the range $|\Delta|>1$ in the current paper, it is thus natural to study the case $|\Delta|<1$ where the distribution of Bethe roots is rather different.

\section*{Acknowledgements}
\label{sec:acknowledgements}
We would like to thank Ivan Kostov and Ben Hoare for helpful comments on the draft. The research of Y.J. is partially supported by the
Swiss National Science Foundation through the NCCR SwissMap.

\appendix
\section{Commutation relations of operators in the XXZ model}
    \label{sec:app-comm-rels}
    We list the commutation relations of the elements of the monodromy matrix, obtained from the $RTT$ relation.
    \begin{align}
        \label{eq:aba-comm-rels}
        \cA(x)\cB(y)&=f(y,x)\cB(y)\cA(x)-g(y,x)\cB(x)\cA(y),\\
        \cB(x)\cA(y)&=f(y,x)\cA(y)\cB(x)-g(y,x)\cA(x)\cB(y)\nonumber, \\
        \cA(x)\cC(y)&=f(x,y)\cC(y)\cA(x)-g(x,y)\cC(x)\cA(y)\nonumber, \\
        \cC(x)\cA(y)&=f(x,y)\cA(y)\cC(x)-g(x,y)\cA(x)\cC(y)\nonumber, \\
        \cB(x)\cD(y)&=f(x,y)\cD(y)\cB(x)-g(x,y)\cD(x)\cB(y)\nonumber, \\
        \cD(x)\cB(y)&=f(x,y)\cB(y)\cD(x)-g(x,y)\cB(x)\cD(y)\nonumber, \\
        \cC(x)\cD(y)&=f(y,x)\cD(y)\cC(x)-g(y,x)\cD(x)\cC(y)\nonumber, \\
        \cD(x)\cC(y)&=f(y,x)\cC(y)\cD(x)-g(y,x)\cC(x)\cD(y)\nonumber, \\
        \left[\cA(x),\cD(y)\right]&=g(x,y)\left(\cC(y)\cB(x)-\cC(x)\cB(y)\right), \nonumber\\
        \left[\cB(x),\cC(y)\right]&=g(x,y)\left(\cD(y)\cA(x)-\cD(x)\cA(y)\right), \\
         \nonumber
        \left[\cA(x),\cA(x)\right] &= \left[\cB(x),\cB(x)\right] = \left[\cC(x),\cC(x)\right] = \left[\cD(x),\cD(x)\right] = 0
    \end{align}
    The functions $f(x,y)$ and $g(x,y)$ are defined as follows
    \begin{equation}
        f(x,y) = \frac{q\frac{x}{y} - q^{-1}\frac{y}{x}}{\frac{x}{y}-\frac{y}{x}}, \quad \quad g(x,y) = \frac{q-q^{-1}}{\frac{x}{y}-\frac{y}{x}}.
    \end{equation}

    \section{Large rapidity expansion}
    \label{sec:app-large-rap}
    In this appendix, we work out the behavior of the $\cB(u)$ and $\cC(u)$ operators in the large rapidity regime, i.e. $u \to \pm \infty$. This will give us the $q$-deformed spin raising and lowering operators. In terms of the multiplicative spectral parameters, these regimes correspond to the limits $x \to 0$ and $x \to \infty$. It turns out to be convenient to slightly twist the $RTT$ relation first, which makes it easier to compute the desired limits.

    \subsection{Gauge transformation}
    The standard $RTT$ relation for the XXZ spin chain reads
    \begin{equation}
        R_{a,b}(x,y)T_a(x)T_b(y)=T_b(y)T_a(x)R_{a,b}(x,y)
    \end{equation}
    with the $R$-matrix as defined in \eqref{eq:aba-r-matrix}, and the monodromy as in \eqref{eq:aba-monodromy}. In order to take the limit of large spectral parameters, we first make the following `gauge transformation':
    \begin{align}
        \tilde{T}_a(x)&= Q(x)T_a(x)Q^{-1}(x)\\
        \tilde{R}_{a,b}(x,y) &= Q(x) \otimes Q(y) R_{a,b}(x,y)Q^{-1}(x)\otimes Q^{-1}(y),
    \end{align}
    where $Q(x)$ is the following matrix acting in the auxiliary space
    \begin{equation}
        Q(x)=\left(
            \begin{array}{cc}
                x^{1/2} & 0 \\
                0 & x^{-1/2}
            \end{array}
        \right).
    \end{equation}
    It is straightforward to see that the $RTT$ relation remains the same under this transformation:
    \begin{equation}
        \tilde{R}_{a,b}(x,y)\tilde{T}_a(x)\tilde{T}_b(y)=\tilde{T}_b(y)\tilde{T}_a(x)\tilde{R}_{a,b}(x,y).
    \end{equation}
    The Lax operator now conveniently decomposes into two triangular matrices:
    \begin{equation}
        \tilde{L}_{a,n}(x)=x L_{a,n}^+-x^{-1} L_{a,n}^-,
    \end{equation}
    where
    \begin{equation}
        L_{a,n}^+ = \left(
            \begin{array}{cc}
                q^{S_n^z} & \left(q-q^{-1}\right)S_n^- \\
                0 & q^{-S_n^z}
            \end{array}
        \right), \quad \quad
        L_{a,n}^- = \left(
            \begin{array}{cc}
                q^{-S_n^z} & 0 \\
                -\left(q-q^{-1}\right) S_n^+ & q^{S_n^z}
            \end{array}
        \right).
    \end{equation}
    It is easy to see that after this transformation, the two limits we want to compute pick out only one of these two terms:
    \begin{equation}
        \lim_{x\to 0} x \tilde{L}_{a,n}(x) = -L_{a,n}^-, \quad \quad \lim_{x \to \infty} \frac{1}{x} \tilde{L}_{a,n}(x) = L_{a,n}^+.
    \end{equation}
    Since the matrices $L^{\pm}_{a,n}$ are triangular, it is now easy to multiply them together in order to obtain the elements of the twisted monodromy matrix. In what follows, we denote these elements as $\tilde{\cA}(x), \tilde{\cB}(x), \tilde{\cC}(x), \tilde{\cD}(x)$. Making use of the twisted $RTT$ relation, we obtain the following commutation relations for the quantum operators:
    \begin{align}
        \tilde{\cA}(x)\tilde{\cB}(y)&=f(y,x)\tilde{\cB}(y)\tilde{\cA}(x)-\frac{y}{x}g(y,x)\tilde{\cB}(x)\tilde{\cA}(y),\\
        \tilde{\cB}(x)\tilde{\cA}(y)&=f(y,x)\tilde{\cA}(y)\tilde{\cB}(x)-\frac{x}{y}g(y,x)\tilde{\cA}(x)\tilde{\cB}(y),\nonumber \\
        \tilde{\cA}(x)\tilde{\cC}(y)&=f(x,y)\tilde{\cC}(y)\tilde{\cA}(x)-\frac{x}{y}g(x,y)\tilde{\cC}(x)\tilde{\cA}(y),\nonumber \\
        \tilde{\cC}(x)\tilde{\cA}(y)&=f(x,y)\tilde{\cA}(y)\tilde{\cC}(x)-\frac{y}{x}g(x,y)\tilde{\cA}(x)\tilde{\cC}(y),\nonumber \\
        \tilde{\cB}(x)\tilde{\cD}(y)&=f(x,y)\tilde{\cD}(y)\tilde{\cB}(x)-\frac{x}{y}g(x,y)\tilde{\cD}(x)\tilde{\cB}(y),\nonumber \\
        \tilde{\cD}(x)\tilde{\cB}(y)&=f(x,y)\tilde{\cB}(y)\tilde{\cD}(x)-\frac{y}{x}g(x,y)\tilde{\cB}(x)\tilde{\cD}(y),\nonumber \\
        \tilde{\cC}(x)\tilde{\cD}(y)&=f(y,x)\tilde{\cD}(y)\tilde{\cC}(x)-\frac{y}{x}g(y,x)\tilde{\cD}(x)\tilde{\cC}(y),\nonumber \\
        \tilde{\cD}(x)\tilde{\cC}(y)&=f(y,x)\tilde{\cC}(y)\tilde{\cD}(x)-\frac{x}{y}g(y,x)\tilde{\cC}(x)\tilde{\cD}(y),\nonumber \\
        \left[\tilde{\cA}(x),\tilde{\cD}(y)\right]&=g(x,y)\left(\frac{y}{x}\tilde{\cC}(y)\tilde{\cB}(x)-\frac{x}{y}\tilde{\cC}(x)\tilde{\cB}(y)\right),\nonumber\\
        \left[\tilde{\cB}(x),\tilde{\cC}(y)\right]&=\frac{x}{y}g(x,y)\left(\tilde{\cD}(y)\tilde{\cA}(x)-\tilde{\cD}(x)\tilde{\cA}(y)\right). \nonumber
    \end{align}
    We see that we recover the original untwisted commutation relations by making the substitutions
    \begin{equation}
        \tilde{\cA}(x) = \cA(x), \quad \tilde{\cB}(x) = x \cB(x), \quad \tilde{\cC}(x) = \frac{1}{x} \cC(x), \quad \tilde{\cD}(x) = \cD(x).
    \end{equation}

    \subsection{Limiting behavior}
We now  compute the operators in the large rapidity regime. From the decomposition of the twisted Lax operator, we find for $x\to\infty$
\begin{equation}
    \lim_{x\to\infty}\frac{\tilde{\cA}(x)}{x^L} = q^{S^z}, \quad \lim_{x \to \infty} \frac{\tilde{\cB}(x)}{x^L} = \left(q-q^{-1}\right)\cS_q^-, \quad\lim_{x\to\infty}\frac{\tilde{\cC}(x)}{x^L} = 0, \quad \lim_{x \to \infty} \frac{\tilde{\cD}(x)}{x^L} = q^{-S^z},
\end{equation}
and for $x \to 0$
\begin{align}
    \lim_{x \to 0} (-x)^L \tilde{\cA}(x) &= q^{-S^z}, \quad \lim_{x \to 0} (-x)^L \tilde{\cB}(x) = 0, \\
    \lim_{x \to 0} (-x)^L \tilde{\cC}(x) &= -\left(q-q^{-1}\right) \cS_q^+, \quad \lim_{x \to 0} (-x)^L \tilde{\cD}(x) = q^{S^z}.
\end{align}
Here we used the operators
\begin{equation}
    q^{\pm S^z} = \prod_{n=1}^L q^{\pm S_n^z}, \quad \cS_q^{\pm} = \sum_{n=1}^L q^{S_1^z} \otimes \cdots \otimes q^{S_{n-1}^z} \otimes S_n^{\pm} \otimes q^{-S_{n+1}^z} \otimes \cdots \otimes q^{-S_L^z}.
\end{equation}
It remains to compute the appropriate limits of the functions $f$ and $g$. For $f(x,y)$ we have
\begin{align}
    \lim_{x\to\infty} f(x,y) &= q, \quad \quad \lim_{y \to \infty} f(x,y) = q^{-1}, \\
    \lim_{x\to 0} f(x,y) &= q^{-1}, \quad \quad \lim_{y \to 0}f(x,y) = q.
\end{align}
For $g(x,y)$ we find
\begin{align}
    \lim_{x\to\infty} \frac{x}{y} g(x,y) &= q-q^{-1}, \quad \quad \lim_{y\to\infty} \frac{x}{y} g(x,y) = 0, \\
    \lim_{x\to 0}\frac{x}{y} g(x,y) &= 0, \quad \quad \lim_{y\to 0}\frac{x}{y} g(x,y) = q-q^{-1},
\end{align}
and
\begin{align}
    \lim_{x\to\infty}\frac{y}{x}g(x,y) &= 0 , \quad \quad \lim_{y\to \infty} \frac{y}{x} g(x,y) = -\left(q-q^{-1}\right), \\
    \lim_{x \to 0} \frac{y}{x} g(x,y) &= -\left(q-q^{-1}\right), \quad \quad \lim_{y \to 0} \frac{y}{x} g(x,y) = 0.
\end{align}

\subsection{Commutation relations}
We can now use our previous results to obtain the commutation relations of the $q$-deformed operators with the original operators. They ultimately read
\begin{align}
    \label{eq:qdef-comm-rels}
    \cS_q^- \tA(x)&= q^{-1} \tA(x) \cS_q^- + \qsz \tB(x), \\
    \cS_q^-\tB(x)&= \tB(x) \cS_q^-, \nonumber\\
    \cS_q^-\tC(x)&=\tC(x) \cS_q^- + \qsz \tD(x)-\qmsz \tA(x), \nonumber \\
    \cS_q^-\tD(x)&= q \tD(x)\cS_q^- - \qmsz \tB(x), \nonumber \\
    \nonumber \\
    \cS_q^+ \tA(x) &= q^{-1} \tA(x) \cS_q^+ - \qmsz \tC(x), \\
    \cS_q^+ \tB(x) &= \tB(x) \cS_q^+ + \qsz \tA(x)-\qmsz \tD(x), \nonumber \\
    \cS_q^+ \tC(x) &= \tC(x) \cS_q^+, \nonumber \\
    \cS_q^+ \tD(x) &= q \tD(x) \cS_q^+ + \qsz \tC(x), \nonumber \\
    \nonumber \\
    \qsz \tA(x) &= \tA(x) \qsz, \\
    \qsz \tB(x) &=q^{-1} \tB(x) \qsz,  \nonumber\\
    \qsz \tC(x) &=q \tC(x) \qsz, \nonumber\\
    \qsz \tD(x) &= \tD(x) \qsz, \nonumber\\
    \nonumber \\
    \qmsz \tA(x) &= \tA(x) \qmsz, \\
    \qmsz \tB(x) &= q \tB(x) \qmsz, \nonumber \\
    \qmsz \tC(x) &=  q^{-1} \tC(x) \qmsz, \nonumber \\
    \qmsz \tD(x) &= \tD(x) \qmsz. \nonumber
\end{align}
If we now perform another large parameter expansion on several of these relations, we indeed obtain the algebra of the quantum group $U_q\left(\mathfrak{sl}_2\right)$:

\begin{equation}
    q^{S^z} \cS_q^{\pm} = q^{\pm 1} \cS_q^{\pm} q^{S^z}, \quad \quad \left[\cS_q^+, \cS_q^-\right] = \frac{q^{2 S^z} - q^{-2S^z}}{q-q^{-1}}.
\end{equation}

\section{Numerical solution of the XXZ Bethe equations}
\label{app:numerics}
\begin{figure}[t]
    \centering
    \includegraphics[width=0.8\linewidth]{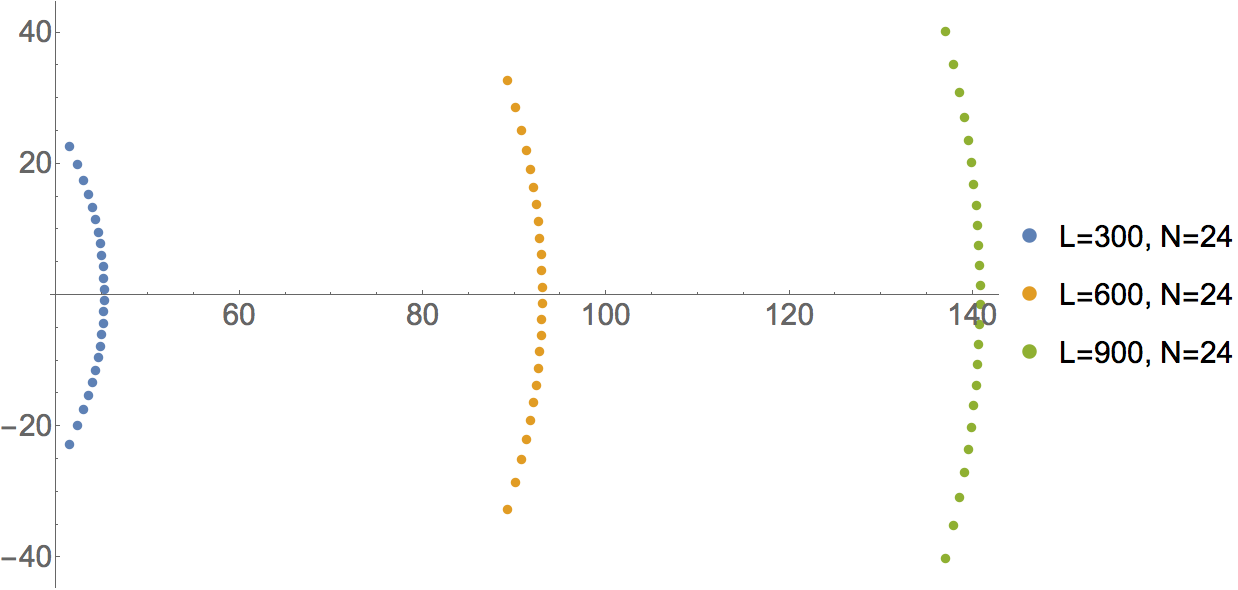}
    \caption{Example of several XXX Bethe root distributions. All sets of roots have mode number $n=1$.}
    \label{fig:app-roots-xxx}
\end{figure}

In this appendix, we provide more detail on how the Bethe root distributions depicted in Figure \ref{fig:contours-xxz} were obtained. As a starting point, we took a set of solutions to the XXX Bethe equations, with parameters $L,N$. When $L$ is  large compared to $N$, these solutions are not hard to find. We apply the \texttt{Mathematica} method \texttt{FindRoot} on the set of of $N$ Bethe equations, where we take the starting point
\begin{equation}
    u_k = \frac{1}{2 \pi n} \left(L+i z_k \sqrt{2L} + \mathcal{O}\left(L^0\right)\right), \quad \quad k=1,\cdots,N,
\end{equation}
where the $z_k$ are the roots of the Hermite polynomial of degree $N$, and $n$ is the mode number. These starting points are sufficiently good for \texttt{Mathematica} to find high-precision numerical solutions to the XXX Bethe equations. An example of resulting sets of roots with several parameters $L,N$ and mode number $n=1$ is shown in Figure \ref{fig:app-roots-xxx}.

With these sets of solutions to the XXX Bethe equations, we can then proceed to obtain solutions to the XXZ equations. We recall these equations for convenience:
\begin{equation}
    \left(\frac{\sinh{\gamma \left(u_i+\frac{i}{2}\right)}}{\sinh{\gamma \left(u_i-\frac{i}{2}\right)}}\right)^L = \prod_{j \neq i}^N \frac{\sinh{\gamma\left(u_i-u_j+i\right)}}{\sinh{\gamma \left(u_i-u_j-i\right)}}, \quad \quad i = 1, \cdots N
\end{equation}
Starting from the isotropic equations (i.e. $\gamma = 0$), we raise $\gamma$ step by step, with small increments (typically 0.001 or 0.0001). We apply the \texttt{FindRoot} procedure at every step, using the result from the previous step as initial guess. We found that this method provides us with fairly accurate sets of solutions, as long as $\gamma$ does not stray too far from zero. As $\gamma$ increases, the exponential aspects of the $\sinh$-functions will give a larger contribution, making the equations more sensitive to small changes in the variables $u_i$. Therefore, the root finding procedure will eventually halt at a certain value of $\gamma$, when it is unable to find a sufficiently accurate set of solutions. It is possible to `push' this boundary a little further by lowering the step size by one or more orders of magnitude, but this turns out to result in fairly insignificant progress.

\begin{figure}[t]
    \centering
    \begin{subfigure}[b]{0.45\linewidth}
        \includegraphics[width=\linewidth]{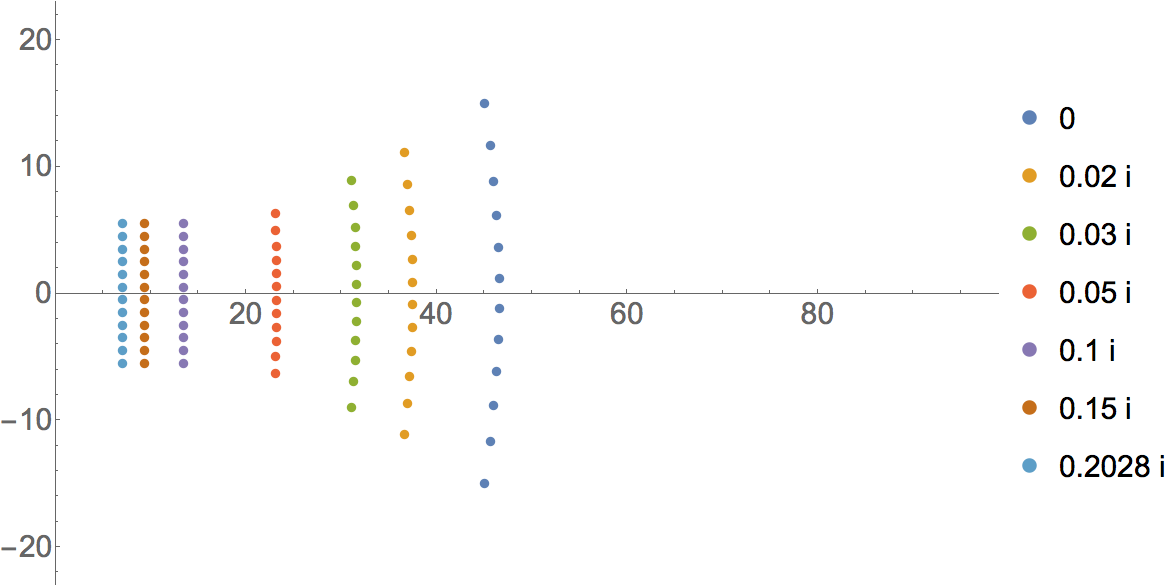}
        \caption{Bethe roots for an $L=300, N=12$ spin chain.}
        \label{fig:app-roots-xxz-300-12}
    \end{subfigure}
    ~~
    \begin{subfigure}[b]{0.45\linewidth}
        \includegraphics[width=\linewidth]{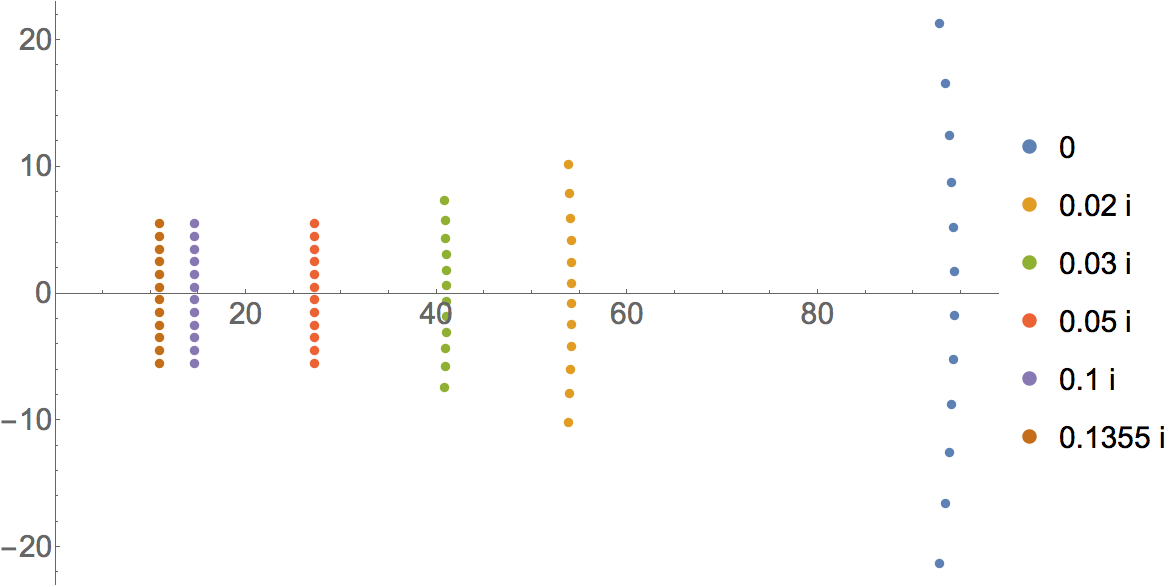}
        \caption{Bethe roots for an $L=600, N=12$ spin chain.}
        \label{fig:app-roots-xxz-600-12}
    \end{subfigure}
    \caption{Comparison of root distributions for two chains of differing length, but equal magnon number. The various sets of roots correspond to different anisotropy parameters $\gamma$, as indicated in the legend.}
    \label{fig:app-comp-length}
\end{figure}

\begin{figure}[t]
    \centering
    \begin{subfigure}[b]{0.45\linewidth}
        \includegraphics[width=\linewidth]{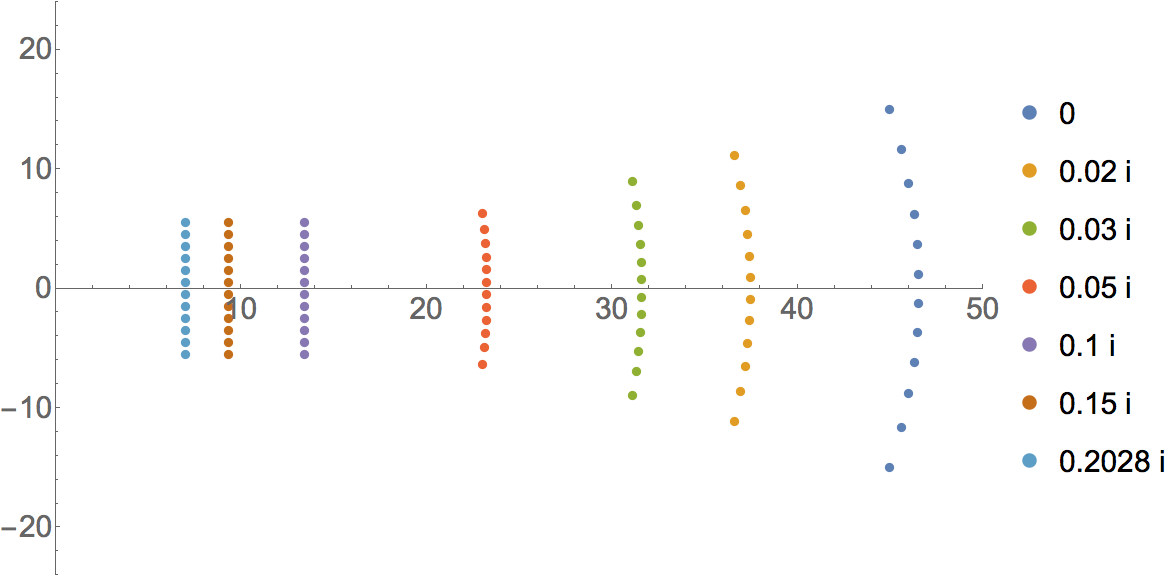}
        \caption{Bethe roots for an $L=300, N=12$ spin chain.}
        \label{fig:app-roots-xxz-300-12-2}
    \end{subfigure}
    ~~
    \begin{subfigure}[b]{0.45\linewidth}
        \includegraphics[width=\linewidth]{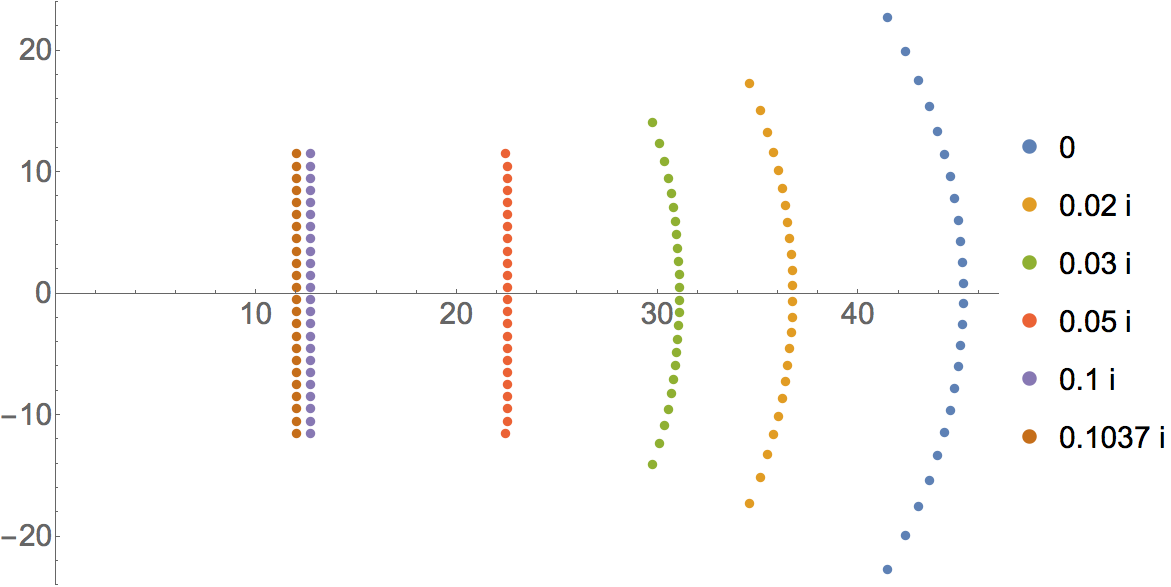}
        \caption{Bethe roots for an $L=300, N=24$ spin chain.}
        \label{fig:app-roots-xxz-300-24}
    \end{subfigure}
    \caption{Comparison of root distributions for two chains of equal length, but different magnon number. The various sets of roots correspond to different anisotropy parameters $\gamma$, as indicated in the legend.}
    \label{fig:app-comp-magnon}
\end{figure}

As mentioned in section \ref{sec:semi-classical-xxz}, the method we used for obtaining the semi-classical limit only applies to the case of a purely imaginary parameter $\gamma$. Therefore, we mainly focused on obtaining root distributions in this regime. As an indication of a typical `halting' value of $\gamma$: for an $L=300, N=12$ spin chain, accurate solutions could be found up to $\gamma=0.2028i$. We illustrate the effect of altering the chain length in Figure \ref{fig:app-comp-length}, and the effect of raising the magnon number in Figure \ref{fig:app-comp-magnon}.

For increasing $L$, it is easy to see that the left-hand side of the equation will be more and more sensitive to variations in the variables $u_i$. Indeed, we find that for longer chains (and similar magnon numbers $N$), the procedure halts at lower values of $\gamma$. For example, keeping the magnon number $N=12$ but doubling the length to $L=600$ resulted in a halting value of $\gamma=0.1355i$. For $L=900$, the procedure already broke down around $\gamma=0.1132i$.

For chains with higher magnon number, the number of equations that have to be solved simultaneously also grows bigger, making the root finding procedure more complicated, and therefore slower. Furthermore, the right-hand side of the Bethe equations becomes a more complicated product, and therefore also more unstable under small changes in the variables $u_i$. We indeed found that increasing the magnon number makes it harder for \texttt{Mathematica} to find good solutions - every step in the iteration takes longer to complete, and the procedure breaks down for smaller values of $\gamma$ compared to chains with less magnons. As an example, we applied the procedure to a chain of length $L=300$ and doubled the magnon number to $N=24$. It then broke down around $\gamma=0.1067i$, a significantly lower value than when we considered only 12 magnons.

\section{Dilogarithm and Quantum dilogarithm}
\label{sec:DL}
In this appendix, we give the definition of Faddeev's quantum dilogarithm function which we use in the main text. Notice that the classical dilogarithm function can be written in the integral form
\begin{align}
\mathrm{Li}_2(-e^{iz})=\frac{i}{2}\int_{\mathbb{R}+i0}\frac{e^{zt}}{\sinh(\pi t)}\frac{dt}{t^2}
\end{align}
where the integral is taken along a line slightly above the real axis. When we close the contour on the upper half plane and use the residue theorem to compute the above integral, we obtain the usual infinite series representation of the dilogarithm function. Replacing $t^2$ in the denominator by $t\sin(b^2t)$ and taking the exponential, we find the definition of Faddeev's quantum dilogarithm\footnote{There is a minor difference of convention between our definition and the standard definition given in the literature, for example \cite{Faddeev2014}.}
\begin{align}
\label{eq:QD}
\Phi_b(z)=\exp\left(\frac{i}{2}\int_{\mathbb{R}+i0}\frac{e^{zt}}{\sin(b^2\,t)\sinh(\pi t)}\frac{dt}{t}\right)
\end{align}
It is easy to see that for $b\to 0$, the quantum dilogarithm reduces to the classical dilogarithm
\begin{align}
\Phi_b(z)=1+\frac{\mathrm{Li}_2(-e^{iz})}{b^2}+\mathcal{O}(1)
\end{align}
Again using the residue theorem, we obtain a series expansion
\begin{align}
\Phi_b(z)=\exp\left(\sum_{n=1}^{\infty}\frac{(-1)^n e^{inz}}{n\,\sinh(b^2\, n)}\right)
\end{align}
It is thus clear that the semi-classical limit of the $q$-deformed $\mathscr{A}$-functional (\ref{eq:semiAXXZ}) for $\gamma=i\phi$ can be written in terms of this function as
\begin{align}
\log\mathscr{A}_{\mathbf{u}}^q[\chi]=\oint_{\mathcal{C}_{\mathbf{u}}}\frac{dz}{2\pi i}\log\Phi_{\sqrt{\phi}}\left(g(z)+\pi\right)
\end{align}

\end{document}